\documentclass[preprint]{aa}
\usepackage{txfonts}
\usepackage{graphicx,subfigure}
\usepackage{epsfig}
\usepackage{epsf}
\usepackage{lscape}
\usepackage{natbib}
\usepackage{rotating}
\usepackage{tabularx}
\newcommand{\mic}{\,{\rm \mu m} }

\begin{document}
\title{Modeling and predicting the shape of the far-infrared to submillimeter emission in ultra-compact HII regions and cold clumps}

\author{D. Paradis \inst{1,2} 
\and C. M\'eny \inst{1,2} \and
 A. Noriega-Crespo \inst{3,6} \and R. Paladini \inst{3} \and
 J.-P. Bernard \inst{1,2} \and C. Bot
 \inst{4} \and L. Cambr\'esy \inst{4}  \and K. Demyk  \inst{1,2} \and  V. Gromov
  \inst{5} \and A. Rivera-Ingraham \inst{1,2} 
 \and M. Veneziani \inst{3} }  
\institute{Universit\'e de Toulouse; UPS-OMP; IRAP; Toulouse, France 
\and
CNRS; IRAP; 9 Av. du Colonel Roche, BP 44346, F-31028, Toulouse, cedex
4, France
\and
Infrared Processing Analysis Center, California Institute of Technology, 1200
E. California Blvd, Pasadena, CA 91125, USA
\and
Observatoire astronomique de Strasbourg, Universit\'e de Strasbourg,
CNRS, UMR 7750, 11 rue de l'Universit\'e, 67000 Strasbourg, France
\and
Space Research Institute, RAS, 84/32 Profsoyuznaya, 117810 Moscow,
Russia
\and
Space Telescope Science Institue, 3700 San Martin Dr, Baltimore, MD,
21218, USA
}

\authorrunning{Paradis et al.}
\titlerunning{Emission from UCHII regions and cold clumps}
\date{}
\abstract
{Dust properties are very likely affected by the environment in which dust
  grains evolve. For instance, some analyses of cold clumps (7 K- 17
  K) indicate that the aggregation process is favored in dense environments. However,
  studying warm (30 K-40 K) dust emission at long wavelength
  ($\lambda$$>$300 $\mic$) has been limited because it is difficult to combine far infrared-to-millimeter (FIR-to-mm) spectral coverage and high
  angular resolution for observations of warm dust grains.} 
{Using Herschel data from 70 to 500 $\mic$, which are part of the Herschel
  infrared Galactic (Hi-GAL)
  survey combined with 1.1 mm data from the Bolocam
  Galactic Plane Survey (BGPS), we compared emission in two types of environments:
  ultra-compact HII (UCHII) regions, and cold molecular clumps (denoted
  as cold clumps). With this comparison we tested dust emission models in the FIR-to-mm domain that reproduce emission in the diffuse medium,  in these two environments
  (UCHII regions and cold clumps). We also investigated their ability
to predict the dust emission in our Galaxy.}
{We determined the emission spectra in twelve UCHII regions and twelve
  cold clumps,
and derived the dust temperature (T) using the recent two-level system
(TLS) model with three sets of parameters and the so-called T-$\beta$
 (temperature-dust emissvity index)
phenomenological models, with $\beta$ set to 1.5, 2 and 2.5.}  
{ We tested the applicability of the TLS model in warm regions for the first time. This analysis indicates distinct trends in the dust emission between
  cold and warm environments that are visible through changes in the dust emissivity
  index. However, with the use of standard parameters, the TLS model is
  able to reproduce the spectral behavior observed in cold and warm
  regions, from the change of the dust temperature alone, whereas a
  T-$\beta$ model requires $\beta$ to be known.} 
{}


\maketitle
\section{Introduction}

The study of the extended far-infrared (FIR) and submillimeter (submm)
sky emission is a relatively young subject. This wavelength range is
dominated by emission from large (15 to 100 nm) silicate-based
interstellar grains (also called big grains, or BG) that dominate the total dust mass and radiate at
thermal equilibrium with the surrounding radiation field. The FIR-to-submm
emission is routinely used to infer total gas column density and mass
of objects ranging from molecular clouds to entire external galaxies,
assuming that dust faithfully traces the gas. Lacking sufficient
observational data in the past century, the emission
was expected to follow the so-called T-$\beta$ model, assuming an optically thin medium and a single
dust temperature along the line of sight,
\begin{equation}
I_{\nu}(\lambda) =\epsilon (\lambda_0) \left ( \frac{\lambda}{\lambda_0} \right )^{-\beta} B_{\nu}(\lambda, T) N_H =Q_{abs}(\lambda) B_{\nu}(\lambda, T)
N_H, 
\end{equation}
with $\beta$=2. $I_{\nu}$ is the sky brigthness, $\epsilon (\lambda_0)$ is the
emissivity at the reference wavelength $\lambda_0$, $B_{\nu}$ is the Planck function, $T$
is the thermal dust temperature, $N_H$ is the hydrogen column density,
and $Q_{abs}$ is the absorption efficiency. T-$\beta$ model with $\beta$=2
is the correct
asymptotic behavior (toward long wavelengths) of the Lorentz
model (the well-known and successful physical model for bound
oscillators). The Lorentz model describes the mid-IR vibrational bands of the
silicate-based interstellar grains. 
\\

Balloon (PRONAOS, Archeops) and satellite (FIRAS, WMAP) mission have
measured the extended interstellar emission in various photometric
FIR, submm, and mm bands. These data analyses have revealed that the
FIR-to-submm emission cannot be explained by a simple extrapolation of
the mid-IR emission. Based on these observations, the FIR-to-submm emission is 
often modeled with the T-$\beta$ model, with $\beta$ taken as a free
parameter, mainly in the
range 1 to 3. With a constant $\beta$  from FIR
to submm and different from 2, this model became an empirical model, that can
consequently hide any possible more complex dependences of the
emissivity with wavelength and temperature.  Two main patterns were observed: \\
- the observed FIR-to-submm dust emissivity ($\epsilon(\lambda)$) appears to
  have a more complex dependence on wavelength than described by the
  T-$\beta$ model: the emission spectrum becomes
flatter in the submillimeter than a modified black-body
emission with $\beta=2$ \citep{Reach95, Finkbeiner99, Galliano05,
  Paladini07, Paradis09, Paradis11}.  This has led to an empirical
change in the optical constants of the Draine astro-silicates
\citep{Draine84} for
wavelengths larger than 250 $\mic$ \citep{Li01}.\\
- the dust emissivity appears to be temperature-dependent in the way that the emissivity spectra are 
flatter with increasing dust temperature \citep{Dupac03,
  Desert08, Veneziani10}. When the dust emission is modeled with the
standard T-$\beta$ model,  a degeneracy between T
and $\beta$ parameters has been highlighted by the various methods of data fitting
($\chi^2$, hierarchical Bayesian, etc ). Therefore, noise can change
the best-fit solution (decreasing T and increasing $\beta$, or vice versa). However, a systematic anti-correlation of $\beta$ with temperature is
claimed to persist \citep{Juvela13}. Similar variations of $\beta$
with temperature have been reported from laboratory
spectroscopic experiments on amorphous dust analogs \citep{Mennella98,
  Boudet05, Coupeaud11}. 

These preliminary results have been confirmed using Herschel
photometric data \citep{Paradis12, Paradis10} as part of
the Hi-GAL survey, an Herschel open time key-project (PI S. Molinari) that mapped the
entire Galactic plane (GP) of our Galaxy \citep{Molinari10a,
  Molinari10b}. In the Herschel wavelength range, dust emissivity spectral variations are often identified with a 500 $\mic$
emission excess \citep{Gordon10, Galliano11, Paradis12}. In the Large
Magellanic Cloud, this excess has been shown to correlate
with temperature and to anti-correlate with brightness
\citep{Galliano11}. A similar behavior is found along the GP using Hi-GAL photometric data where a significant 500 $\mic$
excess is observed toward the peripheral regions of the GP (35$\degr$ $<$ l $<$ 70$\degr$), and can reach up to 16-20$\%$ of the
emissivity \citep[see][Fig. 1, panel A]{Paradis12}. The excess is often highest
($>$25$\%$) toward HII regions, but it does not appear to be 
systematic. However, the Herschel spectral coverage is
limited to 500 $\mic$.  \\

Dust emission and dust processes occurring in warm/hot
environments such as ultra-compact HII (UCHII) regions are poorly known in the FIR-to-submm wavelength range.
 These regions are some of the most luminous objects in the Galaxy at
FIR wavelengths, with dust temperatures of up to 80 K, and are ideal
targets to search for warm/hot dust emission. HII regions correspond
to photoionized regions surrounding O and B stars. UCHII regions are
small (linear size smaller than 0.1 pc) and dense
(electronic density n$\rm _e$ $>$ 10$^4$ cm$^{-}$3), with newly formed O and B stars, before the
ionized gas extends to become compact HII regions. They have
been identified using the IRAS Point Source Catalog (PSC), based on the
[25-12] and [60-12] colors \citep{Wood89a}. UCHII regions have various properties (size,
brightness temperature) and morphologies \citep[cometary, spherical,
core-halo, arc-like, shell, or more complex, see][]{Peeters02}, that
are significantly different
from the standard Stromgren sphere model \citep{Wood89b}
depending on the complex interaction of hot stars and their natal
molecular cloud. The different morphologies of the UCHII regions come
from the ambient medium surrounding the star, but also from the strong
stellar winds of the O and B stars, which create a cavity in the ionized
gas, or from the motion of the star through the cold molecular
gas. All these conditions might  affect dust properties
inside the UCHII regions. Grain destruction/fracturing 
might take place in UCHII regions. In addition, the radiation field
might modify the grain surface, which in turn might change the dust
emissivity. \\
\begin{table}[!t]
\caption[]{Galactic coordinates (in degree) of the selected UCHII regions and cold
  clumps. \label{tab_coord}}
\begin{center}
\begin{tabular}{l c c}
\hline
\hline
Regions & GLON & GLAT \\
\hline
IRAS 17279-3350  & 354.204  & -0.036 \\ 
IRAS 17455-2800 & 1.126 & -0.109 \\ 
IRAS 17577-2320 & 6.554 & -0.098 \\ 
IRAS 18032-2032 & 9.620 & 0.197 \\ 
IRAS 18116-1646 & 13.873 & 0.282 \\
IRAS 18317-0757 & 23.954 & 0.150 \\
IRAS 18434-0242 & 29.955 & -0.014 \\
IRAS 18469-0132 &  31.395 & -0.255 \\
IRAS 18479-0005 & 32.795 & 0.192 \\
IRAS 18502+0051 & 33.914 & 0.109 \\
IRAS 19442+2427 & 60.885 & -0.129 \\
IRAS 19446+2505 & 61.477 & 0.091 \\
\hline
Cold clump 1 & 17.923 & -0.006 \\
Cold clump 2 &17.964 & 0.079 \\
Cold clump 3 & 18.314 & 0.035 \\
Cold clump 4 & 18.104 & 0.379 \\
Cold clump 5 & 18.349 & -0.273 \\
Cold clump 6 & 18.411 & -0.291 \\
Cold clump 7 & 18.572 & -0.431 \\
Cold clump 8 &  18.559 & -0.153 \\
Cold clump 9 & 30.006 & -0.270 \\
Cold clump 10 & 41.715 & 0.035  \\
Cold clump 11 & 42.874 & -0.180 \\
Cold clump 12 & 52.342 & 0.324  \\
\hline
\end{tabular}
\end{center}
\end{table}
In the opposite temperature regime, cold clumps, which are associated
with molecular
clouds, evidence dust emitting at temperatures between $\sim$7 K and $\sim$17
K. Analyses of cold cores allow us to study the initial phases of star
formation, that is the pre-stellar core fragmentation. In the past
(before the Planck and Herschel observations), these objects were poorly
detected in surveys covering wavelengths below 200 $\mic$ because of
their low temperatures and weak emission. Some of them were already studied in the submm and mm domain with ground-based
facilities, however. Recently, $\approx$10000 cold clumps
have been cataloged using Planck data \citep{Planck11}, and some of them were observed
with Herschel in specific programs. Their emission spectra show high dust emissivity index,
sometimes as high as 3.5. This observed behavior might result from
grain coagulation in dense and cold environments \citep{Stepnik03, Paradis09, Kohler11, Kohler12}.\\   

Althgouh FIR-to-mm emission is commonly modeled with a modified black
body, an alternative model has been developped by \citet{Meny07} and
is referred to as the two-level system (TLS) model in the following. It is a physical model for silicate
BG emission in the FIR-to-mm range, derived from solid-state modeling of
general optical properties of the dielectric amorphous state. This
model qualitatively agrees with laboratory experiments on amorphous
silicates, and is coherent with some observational facts, such as the
flattening of the emission at long wavelengths. It is also compatible
with some observations in the amplitude of this flattening in the
Galactic plane observed with Herschel data \citep{Paradis12}. Without
excluding other effects such as the temperature distributions
along the line of sight, grain aggregations, and carbon layers on
silicate-based grains, it is important to interpret the
observations in terms of emission of dielectric grains (silicate maybe
mixed with ice) that radiates at a single temperature along the line of
sight. The observations can be modeled with the four free
parameters of the TLS
model \citep{Paradis11}, allowing us to reproduce the Galactic diffuse medium (denoted in
the following as diffuse parameters), Galactic compact sources
(denoted as compact source parameters), and both environments (denoted
as standard parameters). A full understanding of the observed dust
emission would require a detailed analysis including radiative
transfer, a distribution of grain sizes, a description of the
morphology (aggregation, ice mantles, etc.) of the grains, and the
true IR, FIR, and mm properties of the various materials that are
present in the grain distribution (various silicates, ices and carbon
types, with some control on their degree of amorphisation,
hydrogenation, etc.). Some tests on temperature mixing along the
  line of sight in the inner Galactic plane have been performed in
  \citet{Paradis12}. The authors showed that the
  changes in the observed emissivity spectra with dust temperature
  cannot be accounted for by a line-of-sight effect alone, but might
  instead result from intrinsic variations in the dust properties that
  depend on the environment. \\

\begin{table*}[!t]
\begin{center}
\begin{tabular}{lllllll }  
\hline
\hline
Regions &  $F_{70}$  & $F_{160}$  & $F_{250}$  & $F_{350}$  &$F_{500}$  & $F_{1100}$  \\
\hline
IRAS 17279-3350 (1) & 245.09$\pm$32.74 &
305.20$\pm$37.85  &  154.46$\pm$19.33 & 81.16$\pm$12.12 & 27.09$\pm$6.23  & 0.89$\pm$1.22 \\
IRAS 17279-3350 (2) & 19.24$\pm$7.47 &  16.65$\pm$6.25 &  8.31$\pm$3.65 &
2.72$\pm$1.11 &  0.78$\pm$0.46 & 0.06$\pm$ 0.03 \\
IRAS 17455-2800 (1) & 921.41$\pm$117.10 &  889.23$\pm$98.62 &  362.84$\pm$40.15 &153.04$\pm$21.80 & 48.41$\pm$10.21&
3.43$\pm$2.23 \\
IRAS 17455-2800 (2) & 31.59$\pm$16.34 & 35.18$\pm$15.29 & 26.16$\pm$9.20 & 12.93$\pm$4.96  &  4.30$\pm$1.72 & 0.12$\pm$0.07 \\
IRAS 17577-2320 (1) & 615.98$\pm$68.52 &
408.93$\pm$49.17 & 184.77$\pm$22.82 & 86.23$\pm$13.36 & 29.58$\pm$7.42 & 1.91$\pm$1.81\\
IRAS 17577-2320 (2) & 23.21$\pm$7.66  &  21.64$\pm$7.89 &
12.38$\pm$3.86 &  4.58$\pm$1.53 & 1.91$\pm$0.59 &  0.11$\pm$0.05 \\
IRAS 18032-2032 (1) & 2469.66$\pm$252.86  &
2014.03$\pm$207.58  &   626.05$\pm$52.85  & 300.14$\pm$28.93 &
118.61$\pm$14.65 &  5.97$\pm$2.95 \\
IRAS 18032-2032 (2) & 39.31$\pm$29.51 &  42.27$\pm$16.37 &  22.11$\pm$10.21  & 8.26$\pm$4.07 &  2.44$\pm$0.82 &  0.11$\pm$0.04 \\
IRAS 18116-1646 (1) & 1599.22$\pm$166.58 &
1096.38$\pm$116.61 &  426.07$\pm$39.01 & 180.55$\pm$20.24 &
61.16$\pm$9.92 &  3.50$\pm$2.27 \\
IRAS 18116-1646 (2) &  45.06$\pm$29.29 & 37.45$\pm$12.57   & 16.82$\pm$5.32 &  5.93$\pm$1.70&  1.69$\pm$0.65 &  0.10$\pm$0.04 \\
IRAS 18317-0757 (1) & 1378.92$\pm$144.14 &
759.41$\pm$82.87  &  225.24$\pm$26.13  & 84.00$\pm$14.14  &
25.89$\pm$7.51 & 1.77$\pm$1.78 \\
IRAS 18317-0757 (2) & 30.07$\pm$13.07  &  25.80$\pm$10.02 & 16.98$\pm$6.62 & 6.99$\pm$ 2.72 &  2.39$\pm$1.01 &  0.12$\pm$0.05 \\
IRAS 18434-0242 (1) & 3037.49$\pm$309.69 &
1801.52$\pm$187.72 &  367.84$\pm$37.81 & 229.12$\pm$24.56 &
83.98$\pm$12.58 &  3.59$\pm$2.58 \\
IRAS 18434-0242 (2) & 32.61$\pm$35.43 & 57.02$\pm$27.25 & 30.15$\pm$10.95 &
9.92$\pm$3.74 &   3.96$\pm$1.60 & 0.24$\pm$0.10 \\
IRAS 18469-0132 (1) &  720.85$\pm$78.44  &
646.83$\pm$70.46 & 311.22$\pm$28.98 & 148.69$\pm$16.71 &
53.52$\pm$8.52 &  2.28$\pm$1.72 \\
IRAS 18469-0132 (2) & 18.81$\pm$11.17 & 14.62$\pm$9.55 & 4.69$\pm$2.89 & 2.07$\pm$1.06 &  0.54$\pm$0.28 &  0.05$\pm$0.03 \\
IRAS 18479-0005 (1)& 2361.72$\pm$241.57 &
1581.69$\pm$163.89  & 495.22$\pm$42.90 & 282.81$\pm$26.81 &
92.69$\pm$12.09 &  5.82$\pm$2.79 \\
IRAS 18479-0005 (2) & 17.84$\pm$9.50 &    22.40$\pm$7.15  &   12.66$\pm$5.46  &  4.13$\pm$1.66 &   1.18$\pm$0.55  & 0.06$\pm$0.03 \\
IRAS 18502+0051 (1) & 1096.06$\pm$115.00 &   1076.50$\pm$113.54 &
472.44$\pm$41.11 &   235.61$\pm$23.78   &  80.35$\pm$11.35  &  2.86$\pm$2.11 \\
IRAS 18502+0051 (2) & 9.97$\pm$5.38  & 19.59$\pm$ 6.63 & 11.22$\pm$4.75 &  5.32$\pm$2.03&    1.66$\pm$0.65 &   0.11$\pm$0.04 \\
IRAS 19442+2427 (1) & 1526.10$\pm$158.99 &
874.65$\pm$95.40 &  434.83$\pm$39.89 & 190.44$\pm$21.63 &  72.86$\pm$11.18  & 3.96$\pm$2.52 \\
IRAS 19442+2427 (2) & 38.06$\pm$14.83 &  47.92$\pm$18.29 &  19.69$\pm$7.66 &
8.57$\pm$3.61 &   2.34$\pm$ 0.95 &  0.15$\pm$0.08 \\
IRAS 19446+2505 (1) & 3851.74$\pm$392.29  &  1716.05$\pm$179.22 &
553.55$\pm$50.20 &  236.73$\pm$24.90 &  79.65$\pm$11.87 &
5.51$\pm$2.96 \\
IRAS 19446+2505 (2) & 134.00$\pm$79.70 &   70.81$\pm$35.01 &  25.71$\pm$10.49 &
8.00$\pm$3.13  & 2.87$\pm$0.77 &  0.17$\pm$0.08 \\        
\hline
Cold clump 1 (1) & 7.73$\pm$2.95 &
13.42$\pm$4.30 & 10.40$\pm$3.58 & 4.65$\pm$2.33 &
2.18$\pm$1.53 & 0.14$\pm$0.37 \\
Cold clump 1 (2)  &  0.31$\pm$0.15 & 0.98$\pm$0.45 &
0.83$\pm$0.34 & 0.47$\pm$0.20 & 0.21$\pm$0.10  & 0.02$\pm$0.01 \\
Cold clump 2 (1) & 79.94$\pm$12.23 & 107.14$\pm$15.76 &
68.61$\pm$10.87 & 31.30$\pm$6.94  &  13.17$\pm$4.37 &   0.81$\pm$1.01 \\
Cold clump 2 (2)  &  0.37$\pm$0.31  & 1.47$\pm$0.77 & 1.55$\pm$0.63 &
0.73$\pm$0.28 & 0.34$\pm$0.14 & 0.02$\pm$0.01  \\    
Cold clump 3 (1) & - &  8.89$\pm$5.52 &
8.16$\pm$4.70 &   4.64$\pm$3.19 & 1.82$\pm$2.08 &  0.09$\pm$0.20 \\
Cold clump 3 (2) & - &   1.96$\pm$0.34 &  1.64$\pm$0.40 &
0.83$\pm$0.23 & 0.36$\pm$0.11 &  0.01$\pm$0.01 \\
Cold clump 4 (1) &  5.17$\pm$2.60 & 9.94$\pm$3.41
& 11.35$\pm$3.69  &  7.10$\pm$2.98 &  3.57$\pm$1.96 &
0.21$\pm$0.46 \\
Cold clump 4 (2) & 0.20$\pm$0.08 & 1.28$\pm$0.24  & 1.30$\pm$0.33 &
0.79$\pm$0.31 & 0.27$\pm$0.12 &   0.02$\pm$0.01 \\
Cold clump 5 (1) & 4.19$\pm$3.66 & 19.28$\pm$8.77 &
25.30$\pm$9.07&  17.09$\pm$6.69 & 7.68$\pm$4.30 &
0.39$\pm$0.84 \\
Cold clump 5 (2) &  0.50$\pm$0.11 & 2.79$\pm$1.30& 3.45$\pm$1.52 &
2.00$\pm$0.90 &  0.88$\pm$0.39  &  0.04$\pm$0.02 \\
Cold clump 6 (1) & - &  5.57$\pm$4.65 &
16.25$\pm$6.76 & 14.30$\pm$5.54  & 7.50$\pm$3.78 &
0.21$\pm$0.63  \\
Cold clump 6 (2) & - &  1.18$\pm$0.49 & 1.77$\pm$0.85 & 1.01$\pm$0.66 &
0.46$\pm$0.32 &  0.03$\pm$0.02  \\
Cold clump 7 (1) & 4.49$\pm$2.59 & 4.65$\pm$2.95
& 4.53$\pm$2.40 &  4.71$\pm$2.46 & 2.17$\pm$1.55 &
0.11$\pm$0.33 \\
Cold clump 7 (2) &  0.15$\pm$0.35 & 2.72$\pm$0.55 &
2.06$\pm$0.30  & 0.96$\pm$0.28 & 0.39$\pm$0.11 & 0.02$\pm$0.01 \\
Cold clump 8 (1) &  - &  2.48$\pm$5.75 &
9.00$\pm$6.41 &  6.62$\pm$4.45 & 3.12$\pm$2.66 & 0.13$\pm$0.61 \\
Cold clump 8 (2) & - & 1.80$\pm$0.36 &  2.22$\pm$0.56 &
1.33$\pm$0.32 &   0.58$\pm$0.15 &  0.03$\pm$0.01 \\
Cold clump 9 (1) & 104.97$\pm$14.96 & 189.10$\pm$25.65 &
139.88$\pm$17.64 & 74.53$\pm$11.28 & 31.25$\pm$6.40 &
1.65$\pm$1.33 \\
Cold clump 9 (2) & 0.64$\pm$0.72 & 6.68$\pm$2.66 &
6.23$\pm$2.17 & 3.59$\pm$1.26 & 1.82$\pm$0.56 & 0.07$\pm$0.03 \\
Cold clump 10 (1) & 2.71$\pm$1.86 &
10.80$\pm$3.74  &  8.25$\pm$3.14 & 3.16$\pm$1.87 &
1.45$\pm$1.22 &  0.15$\pm$0.39 \\
Cold clump 10 (2) & 0.88$\pm$0.22 & 2.39$\pm$0.42 & 1.63$\pm$0.30 &
0.80$\pm$0.14 & 0.31$\pm$0.06 &  0.02$\pm$0.01 \\
Cold clump 11 (1) & 0.66$\pm$0.70 &   3.61$\pm$3.59 &
4.14$\pm$4.67 &  3.44$\pm$3.15 & 1.56$\pm$1.91 &
0.08$\pm$0.38 \\
Cold clump 11 (2) & - &  0.75$\pm$0.21 &  0.97$\pm$0.39 &
0.42$\pm$0.16 &  0.20$\pm$0.09 & 0.01$\pm$0.01 \\
Cold clump 12 (1) &  7.57$\pm$2.86 &
21.31$\pm$5.20  &  21.78$\pm$5.13  & 13.39$\pm$3.93 &
7.19$\pm$2.76 &  0.30$\pm$0.55  \\
Cold clump 12 (2) & 0.02$\pm$0.04 &  0.65$\pm$0.28 & 0.95$\pm$0.37  &
0.63$\pm$0.27 & 0.28$\pm$0.11 & 0.02$\pm$0.01 \\
\hline
\end{tabular}
\end{center}
\caption{Fluxes (in Jy) computed in the central and
  surrounding part of the regions, denoted as (1) and (2). \label{tab_fluxes}}
\end{table*}

Most studies in the FIR-to-mm domain require a realistic
determination of dust temperature and dust column density over large
parts of the sky. This is of primary importance for predicting the emission intensity at
any other FIR-to-mm wavelengths, for determining masses, for removing some
Galactic foreground components from cosmological signals (such as
cosmic microwave background), which requires a very accurate
extrapolation in frequency, or for determining variations in the general dust
emission properties in various environments. Therefore, understanding variations in dust
emissivity is crucial. The aim of this work is to compare the
  shape of the dust emission in
different environments to investigate wheter distinct properties can be
distinguished, and if so, to be able to accurately reproduce the
  shape of dust
emission spectra in connection with the environment. However, we also
wish to be able to easily predict dust emission in any regions of our
Galaxy, even when the characteristics of the region, that is types of
the environmment (diffuse, cold, warm), for instance, are unknown. 
We compare the ability of the TLS model with the three sets of
parameters (diffuse, compact sources, standard), and a T-$\beta$ model with three fixed values of $\beta$
(1.5, 2, and 2.5) to fit the BG emission in warm and cold regions of
the Galactic interstellar medium. 
\begin{table*}[!t] 
\caption[]{Best-fit parameters of the TLS model derived for different environments. \label{tab_tls_param}} \begin{center}
\begin{tabular}{l c c c c }
 \hline
\hline
 Environment$/$parameters &  $l_c$ (nm) & $A$ & $c_{\Delta}$ & reduced $\chi^2$ \\ 
 \hline
 Galactic diffuse$^*$: &&&& \\
\,\,\,\,\,\,\,\,\, Diffuse parameters & 23.05 $\pm $ 22.70  &9.38 $\pm $ 1.38 &
242 $\pm$ 123 & 1.95 \\
\hline
 Galactic compact sources$^*$: &&&& \\ 
\,\,\,\,\,\,\,\,\, Compact source parameters & 5.11 $\pm$ 0.09 & 3.86 $\pm$ 0.13
& 1333 $\pm $ 68  & 1.45 \\
\hline
Galactic diffuse and compact sources$^*$:  &&&& \\
\,\,\,\,\,\,\,\,\, Standard parameters & 13.40 $ \pm $ 1.49 & 5.81 $\pm $ 0.09 & 475 $\pm
$ 20 & 2.53 \\ 
\hline
UCHII regions$^{\dag}$: &&&&\\ 
\,\,\,\,\,\,\,\,\, Diffuse parameters & 23.05 $\pm $ 22.70  &9.38 $\pm $ 1.38 &
242 $\pm$ 123 & 1.27 \\
\,\,\,\,\,\,\,\,\, Galactic compact sources & 5.11 $\pm$ 0.09 & 3.86 $\pm$ 0.13
& 1333 $\pm $ 68  & 2.09 \\
\,\,\,\,\,\,\,\,\, Standard parameters & 13.40 $ \pm $ 1.49 & 5.81 $\pm $ 0.09 & 475 $\pm
$ 20 &1.28 \\
 \hline
\end{tabular}
\end{center}
$^*$ \citet{Paradis11}.\\
$^\dag$ This work.\\
\end{table*}

The main goal of this work is to investigate the potentially
  distinct dust properties depending on the environment and to be able
to predict the FIR-to-mm emission in cold and warm regions. 
In this study, we combine Bolocam with Herschel data to extend the
spectral coverage to mm wavelengths (1.1 mm), which is important to detect any
changes in the shape of the emission spectrum. Data from the Midcourse Space
Experiment (MSX) in band E
(21.3 $\mic$) and
Spitzer data at 24 $\mic$ are also presented, but were not included in the
modeling. In sect.
\ref{sec_data} we briefly summarize surveys, in Sect. \ref{sec_env},
we explain the selection of targets in the two specific environments
(UCHII regions and cold clumps). We describe the method
(including dust emission extraction and modeling) in Sect.
\ref{sec_method}. Discussions and conclusions are provided in Sect.
\ref{sec_discussion} and \ref{sec_cl}.

\section{Data}
\label{sec_data}
\subsection{Hi-GAL survey}

The Hi-GAL survey covers the entire Galactic plane (-1$\degr$
$<$b $<$+1$\degr$) at five wavelengths (70, 160, 250, 350 and 500
$\mic$), with an angular resolution going from 6$^{\prime \prime}$ to
37$^{\prime \prime}$. The data was processed with the software ROMAGAL \citep[][]{Traficante11}. The PACS and SPIRE
absolute zero level were calibrated by applying gains and offsets derived from the
comparison with the Planck-High Frequency Instrument and IRIS
\citep[Improved Reprocessing of the IRAS Survey, see][]{MamD05} data
\citep[see][]{Bernard10,Paradis12}. 

\subsection{BGPS survey }

With an angular resolution of 33$^{\prime \prime}$, the Bolocam Galactic
Plane Survey \citep[BGPS, ][]{Aguirre11} covers the
longitude and latitude ranges -10.5$\degr$ $\le$ l $\le$ 90.5$\degr$ and $\mid$b$\mid$$\le$0.5$\degr$ in a contiguous
way. Extentions in latitude were performed in some regions (Cygnus
X spiral arm, l=3$\degr$, 15$\degr$, 30$\degr$ and 31$\degr$). Four
regions in the outer Galaxy were also observed: IC1396, a region toward the Perseus arm,
W3/4/5, and Gem OB1. The total coverage area is 170 square
degrees. A full description of the BGPS can be found in
\citet{Aguirre11}. The data in unit of Jy/beam, were first converted into
MJy/sr using Eq. 16 from \citet{Aguirre11}, which was derived from the beam
surface value. We used the new
version of the data (v2.0, 2013). In this
version, data no longer suffer from calibration issues, that is the
1.5 factor needed by \citet{Aguirre11} in the previous version of
the data to obtain consistency with other data sets has been
made redundant. However, the processing of the maps possibly attenuates 
the aperture flux for structures extending to 3.8$^{\prime}$ by 50$\%$ . 

\subsection{Additional data}

We also analyzed near-infrared (NIR) data such as the MSX data in band E (21.3 $\mic$, with a resolution
of 20$^{\prime \prime}$) and
Spitzer data (24 $\mic$, with a resolution of 6$^{\prime \prime}$) as
part of the MIPSGAL program \citep[PI: S. Carey, ][]{Carey09} for the UCHII
regions, but they were not included in the modeling (see Sect. \ref{sec_modelling}). Most of the UCHII regions
we are interested in are very bright, and some pixels of the 24 $\mic$
images were saturated. These pixels were replaced using MSX band E
data at a lower resolution than the original Spitzer data, which might
result in
underestimated flux. The corrected 24 $\mic$ images are not yet published.  \\

All the data were
convolved to a 37$^{\prime \prime}$ angular resolution to match the
resolution of the Herschel 500 $\mic$ data, with a pixel size of
13.9$^{\prime \prime}$. The resolution was changed by convoluting by a Gaussian kernel with FWHM
  $\sigma_k^2=\sigma_c^2-\sigma_d^2$, where $\sigma_c$ is the common
  resolution, that is 37$^{\prime \prime}$, and $\sigma_d$ is the original
resolution of the data. The SPIRE 500 $\mic$ beam profile has a plateau at
  $\sim1\%$ that extends to a radial distance of 1$^{\prime}$.  The
  Gaussian approximation of the beam is still valid even for
  the selected annulus we consider in the following (28$^{\prime
    \prime}$ to 56$^{\prime}$, see Sect. \ref{sec_aper}).

To avoid any zero level mismatch between Herschel, Spitzer, MSX and Bolocam data,
we subtracted a background from all images. The background
was computed as the median over a common area, corresponding to the
10$\%$ lowest values in the Bolocam data.

\section{Two specific environments}
\label{sec_env}
\subsection{ UCHII regions}
\begin{figure*}[!t]
\begin{center}
\includegraphics[width=16cm]{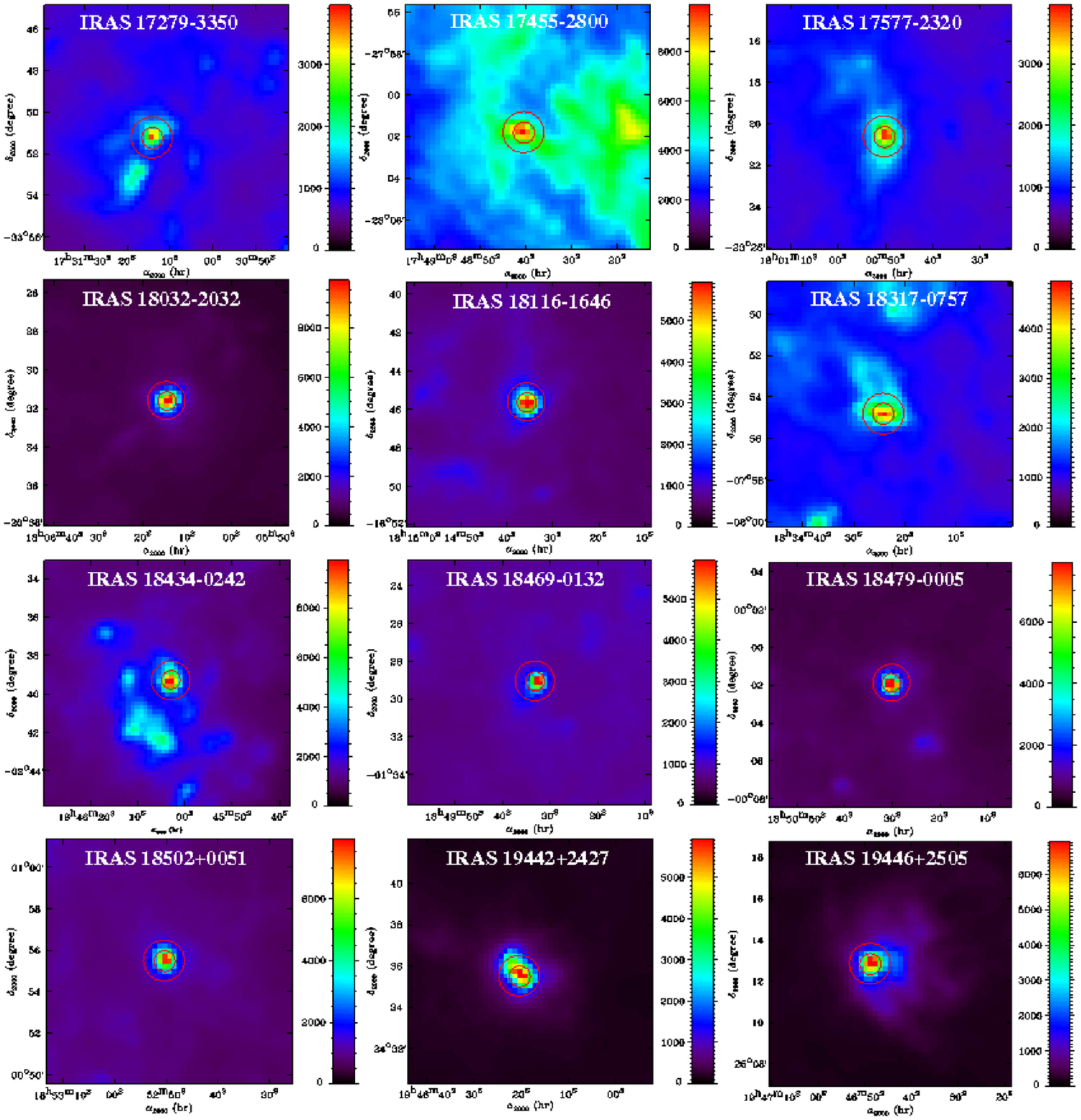}
\caption{Herschel 350 $\mic$ images of the selected UCHII
  regions. The photometric apertures for which the fluxes have been
  computed are indicated by red circles. Central and surrounding
  regions are denoted (1) and (2) in the text. \label{fig_uchii} }
\end{center}
\end{figure*}
The UCHII regions have been cataloged by \citet{Codella94} using the
association of HII regions and IRAS PSC. We chose twelve targets from the
catalog that were observed in both the Hi-GAL and BGPS surveys and have high 100
$\mic$ IRAS fluxes ($> 10^3$ Jy) to ensure that we studied UCHII regions
that include warm dust. Because IRAS has a lower resolution than the Herschel data, the coordinates of the regions were
determined from the maximum surface brightness at 160 $\mic$. Characteristics
and images of the selected UCHII regions are given in Tab. \ref{tab_coord}
and Fig. \ref{fig_uchii}. 

\subsection{Cold clumps}
\begin{figure*}[!t]
\begin{center}
\includegraphics[width=16cm]{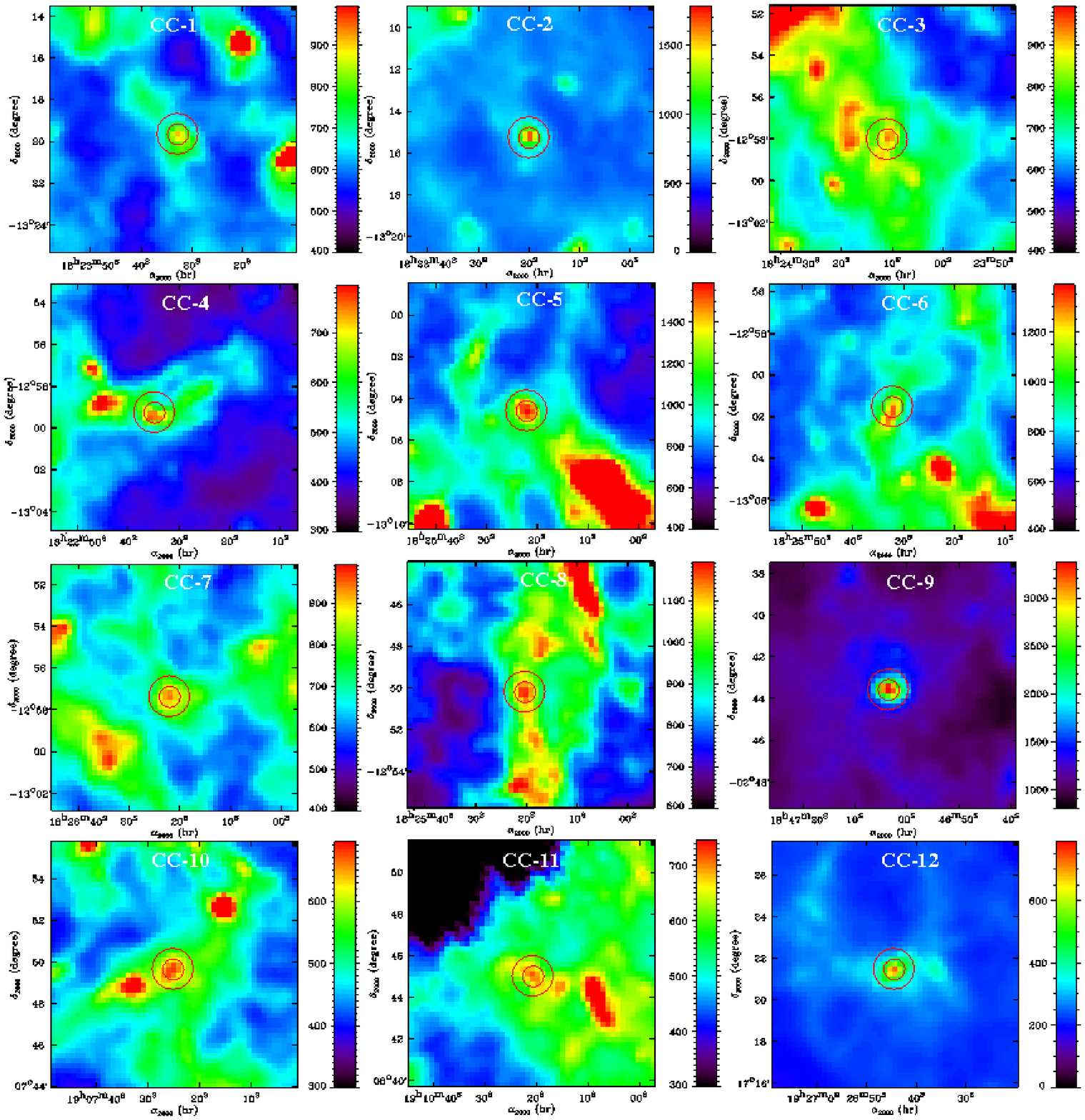}
\caption{Herschel 350 $\mic$ images of the selected cold clump regions. The photometric apertures for which the fluxes have been
  computed are indicated by red circles. Central and surrounding regions are denoted
  (1) and (2) in the text. \label{fig_cc}}
\end{center}
\end{figure*}
We chose cold molecular clumps (previously identified from $^{13}$CO
\citep[observations using the BU-FCRAO Galactic Ring Survey, see ][]{Jackson06},
that were recently analyzed using a 3D - Galactic inversion on Herschel
observations \citep[Tab. 1 in][]{Marshall13}, based on HI and
$^{13}$CO data. In this analysis, dust
temperatures in each phase of the gas have been determined for each
molecular clump. We selected twelve targets that
show cold dust. In the following, we refer to these regions as cold
clumps, even if they do not strictly correspond to the definition 
adopted by the Planck collaboration. For each cloud we obtained the
exact coordinates that enabled us to derive the maximum surface brigthness
at 500 $\mic$ (FIR-to-submm emission peaks do not correspond to HI or
$^{13}$CO peaks). This selection leads to coordinates different from those
reported in \citet{Marshall13}. The coordinates of our cold clump selection are provided in
Tab. \ref{tab_coord}. Images of the targets at 350 $\mic$ are provided
in Fig. \ref{fig_cc}.
\begin{figure*}[!t]
\begin{center}
\includegraphics[width=16cm]{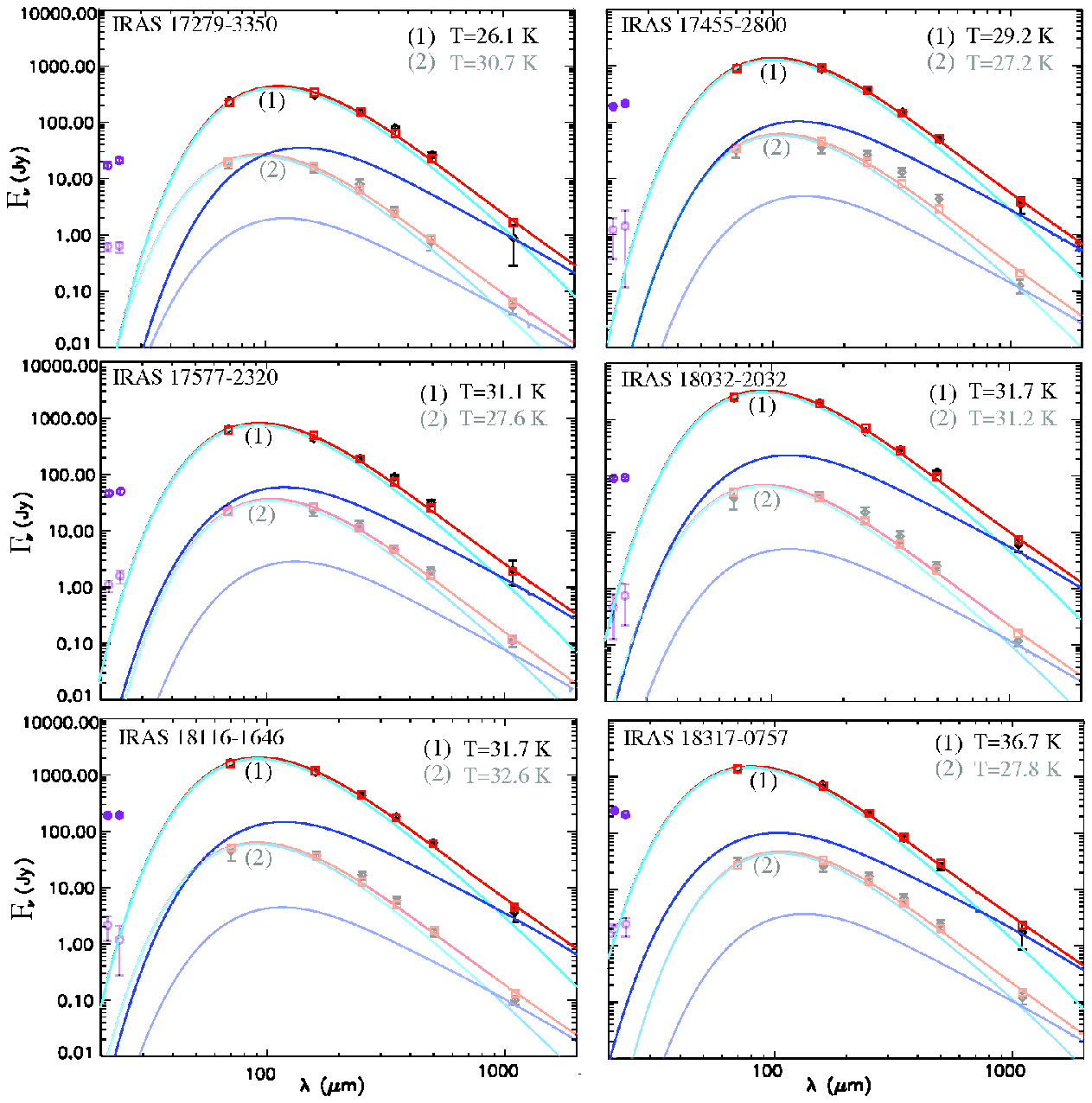}
\caption{Herschel (70, 160, 250, 350, 500 $\mic$) - Bolocam (1.1 mm)
  SED (black diamonds) of UCHII regions, fitted with the TLS model using
  the diffuse parameters (total emission in red, DCD and TLS processes in light
 and dark blue). Squares represent models integrated in
  the band filters of each instrument, allowing direct comparisons
  with data (diamonds). SED corresponding to the central (1)
  and surrounding (2) part of the region are represented in dark and
  light colors. MSX data in band E (21.3 $\mic$) and
  Spitzer data at 24 $\mic$ are also visible in the plots, represented
  by the purple circles. 
\label{fig_uchii_sed}}
\end{center}
\end{figure*}
\addtocounter{figure}{-1}
\begin{figure*}
\addtocounter{subfigure}{1}
\begin{center}
\subfigure{\includegraphics[width=16cm]{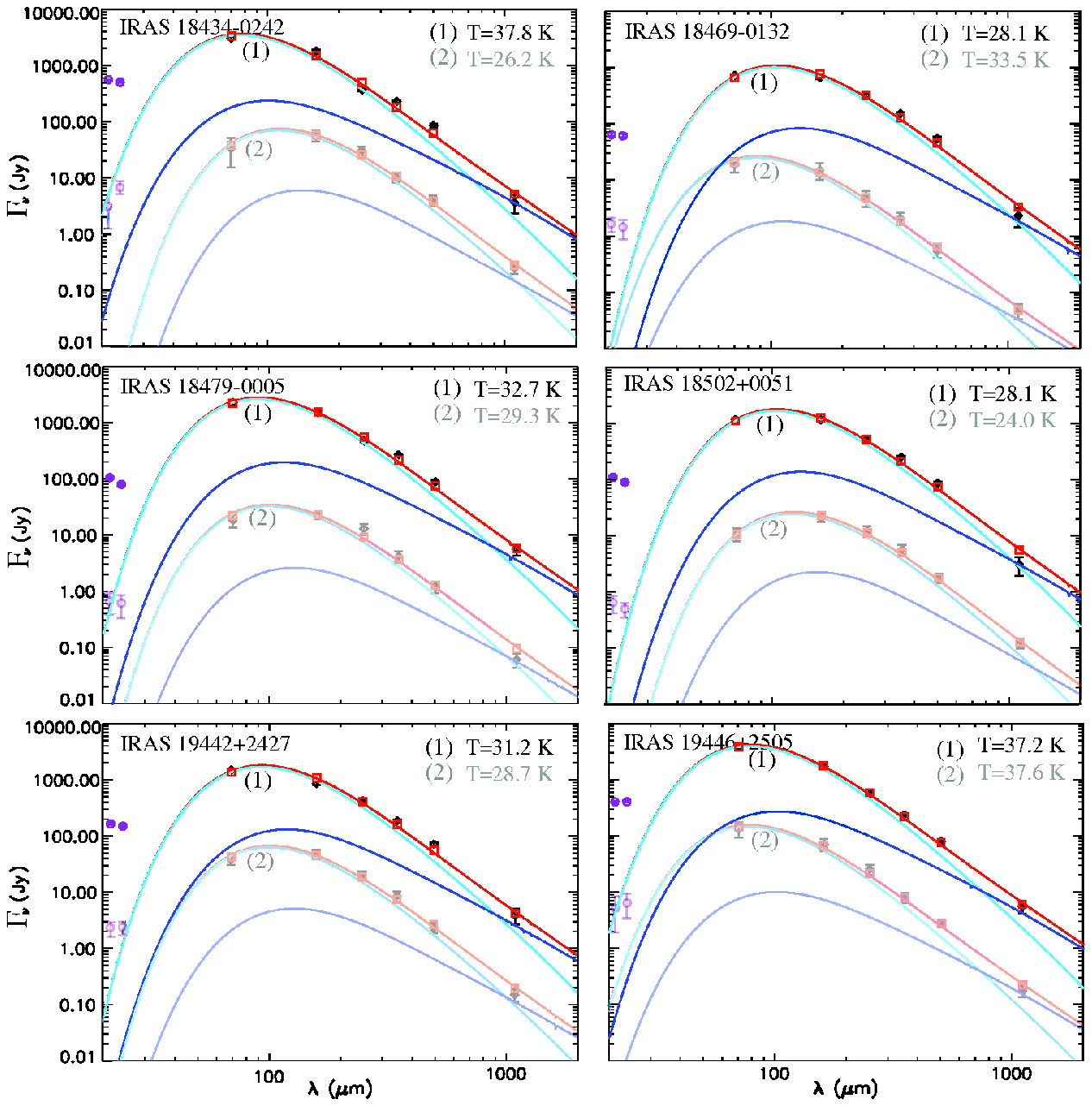}}
\end{center}
\caption{Continued.}
\end{figure*}

\section{Method}
\label{sec_method}
\subsection{Aperture photometry to extract dust emission}
\label{sec_aper}
For each UCHII region and cold clump, we extracted two
spectral energy distributions (SEDs). It is more reasonable to
base our analysis on two SEDs per region than on only one.  Instead of determing SEDs per
pixels, we constructed SEDs derived from averaging several pixels. We
chose the central part of the region
(denoted as -(1)- in the following) that has bright pixels (not
intended to describe the core of the region), and an annulus
surrounding the central part (denoted as -(2)- in the following). In
this way, we expected to obtain slight changes in dust temperature
farther away from the central part, to sample various
temperatures. 
For this purpose we used the idl routine {\it aper} to compute
concentric aperture photometry. We fixed the first aperture to a two-pixel radius (27.8$^{\prime \prime}$) and the surrounding annulus with an inner and outer
radius of two and four pixels (between 27.8$^{\prime \prime}$ and
55.6$^{\prime \prime}$). The SED in region (1) was
background subtracted from the annulus region (2). We considered as
  uncertainty the quadratic sum of the uncertainty deduced from  the
  idl routine {\it aper}, which includes the dispersion on the sky
  background (corresponding to the root mean square of the background), and the calibration uncertainty
depending on each instrument. The Hi-GAL data have been generated by
the software ROMAGAL \citep[][]{Traficante11}, which does not remove
the large-scale emission, as opposed to standard high-pass
filtering. For these data, the calibration uncertainty has been
estimated to be 10$\%$ for PACS \citep{Poglitsch10} and 7$\%$ for
SPIRE (oberver's manual v2.4). For the Bolocam
data, we used a calibration uncertainty of 20$\%$, which corresponds to
the comparison of the v2 version of the data with the flux from other
instruments \citep{Ginsburg13}.  We note that the Bolocam data uncertainties on the output flux from
the routine {\it aper} are large because of noise in the
data. Adding a calibration
uncertainty of 20$\%$, we obtained in some cases a total uncertainty twice (or
even more) larger than the flux, which indicates a low signal-to-noise ratio.
We determined the flux at each
wavelength from 70 $\mic$ to 1.1 mm to obtain FIR-mm SEDs. Flux
values are provided in Tab. \ref{tab_fluxes}. We proceeded in the same
way to also deduce the 21.3 $\mic$ and 24 $\mic$ fluxes in UCHII regions. 
The SEDs are given in Fig. \ref{fig_uchii_sed} and \ref{fig_cc_sed}.
\begin{figure*}[!t]
\begin{center}
\includegraphics[width=16cm]{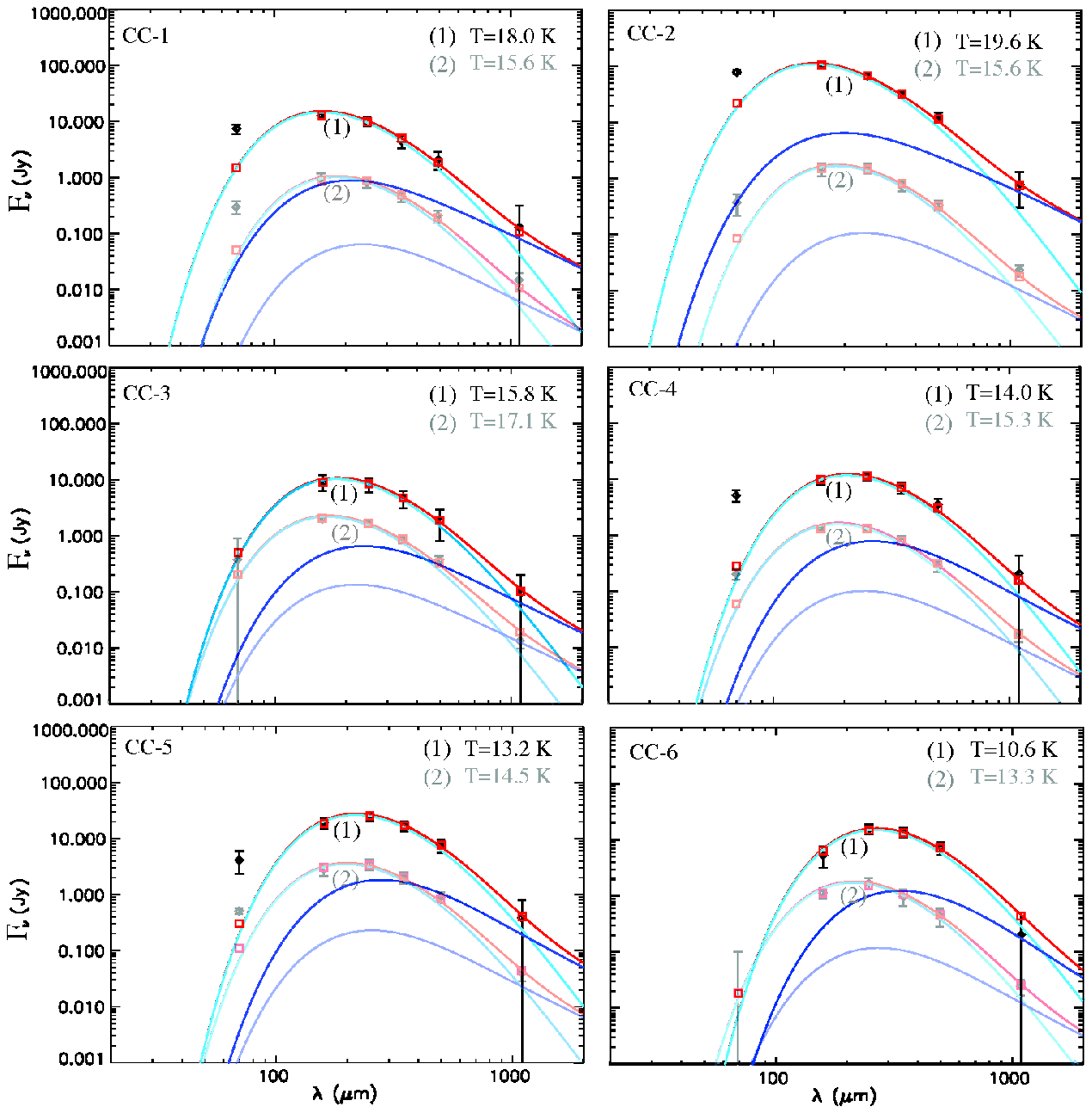}
\caption{Herschel (70, 160, 250, 350, 500 $\mic$) - Bolocam (1.1 mm)
  SED (black diamonds) of cold clumps, fitted with the TLS model using
  the compact source parameters (total emission in red, DCD and TLS processes in light
 and dark blue). Squares represent models integrated in
  the band filters of each instrument, allowing direct comparisons
  with data (diamonds). 70 $\mic$ data have not been
 included in the fitting. SED corresponding to the central (1)
  and surrounding (2) part of the region are represented in dark and
  light colors. \label{fig_cc_sed}}
\end{center}
\end{figure*}

\addtocounter{figure}{-1}
\begin{figure*}
\addtocounter{subfigure}{1}
\begin{center}
\subfigure{\includegraphics[width=16cm]{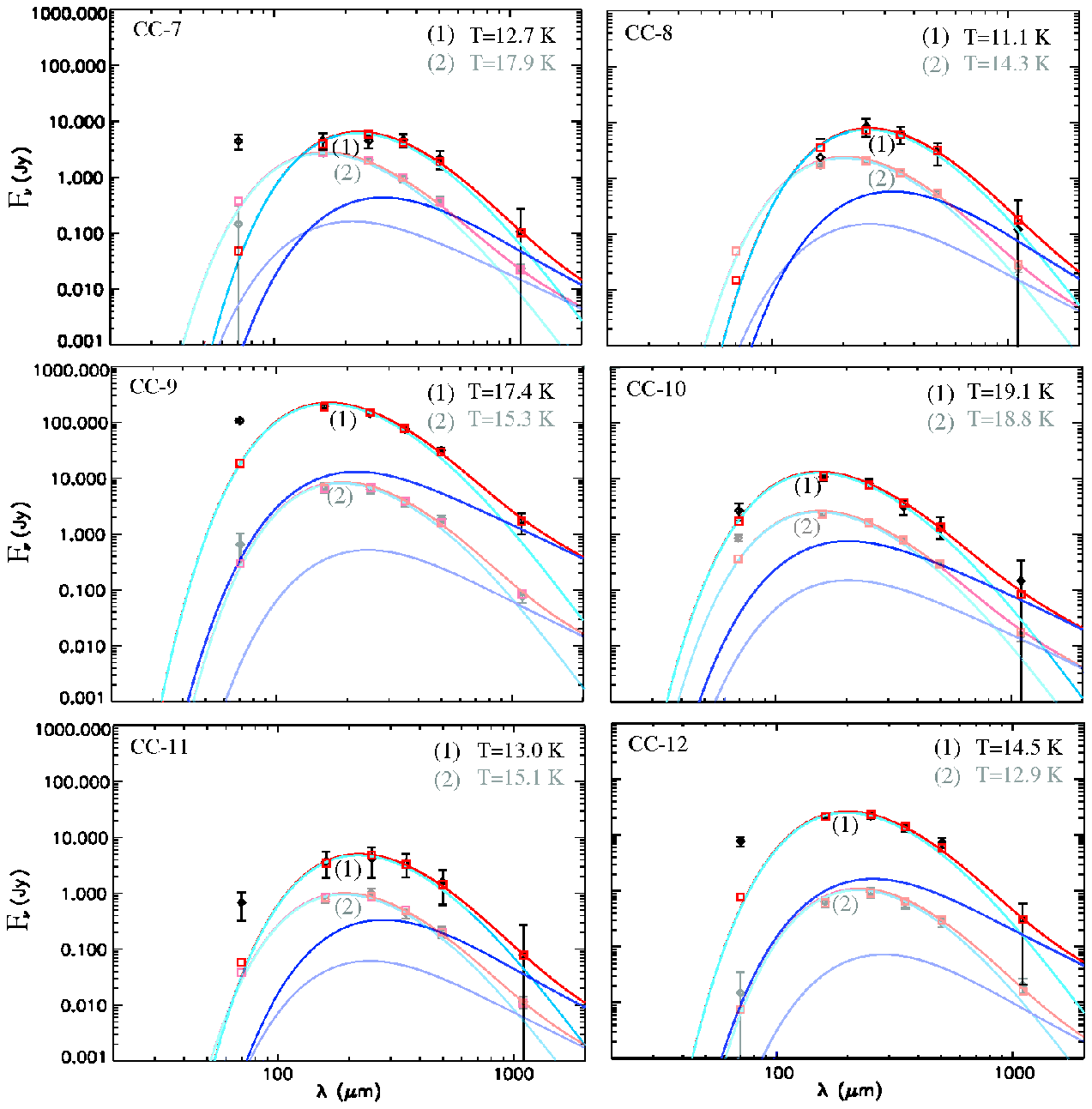}}
\end{center}
\caption{Continued.}
\end{figure*}

\subsection{Modeling}
\label{sec_modelling}
Figure \ref{fig_uchii_sed} shows that fluxes in the NIR
wavelengths are quite high, probably because of the contributon
from small grains that are stochastically heated by the radiation field. Since the models used here
include a single BG component, we selected a wavelength range
where this component clearly dominates the overall emission. In
addition, the wavelength range was restricted by the validity of both
the TLS and the T-$\beta$ models. These models are only valid in the
FIR-to-mm for wavelengths longer than $\sim$50 $\mic$, where the
following assumptions can be made: the real part of the dielectric
constant can be considered to be constant and the size of the particules
can be considered to be smaller than the wavelength. Moreover, the 21
$\mic$ flux in cold clumps might be biased by the absorption resulting
from the silicate bands occuring at 20 $\mic$.Therefore, the 21.3
$\mic$ and 24 $\mic$ flux were not included in the modeling. 
\subsubsection{T-$\beta$ model}
The typical way of describing FIR emission is to use a simple
modified black-body model with a fixed value of $\beta$. The common
value of $\beta$ is 2. This type of model is acceptable when long
wavelength constraints are not available and
for regions with temperatures of about 17-20 K.
However, there is no reason a unique value of $\beta$ to be applicable
throughout the sky. Some authors have claimed that $\beta$ variations
are only a result
of calibration uncertainties on the data, temperature mixing along
the line of sight \citep{Shetty09}, or applications of the
$\chi^2$ minimization technique. A Bayesian approach
on the data modeling, however, can clearly distinguish between a
real and spurious T-$\beta$ relationship \citep{Kelly12,
  Veneziani13}. Moreover, in some cases it is obvious that a modified black-body model with $\beta$=2
 does not work, especially in cold regions with steep
 spectra \citep[$\beta$ $\simeq 3$, see, for instance,][]{Desert08,
     Planck11}, or hot regions with flat spectra \citep[$\beta$
   $\simeq 1$, see, for instance,][]{Dupac03, Kiuchi04}. However, this is not a systematic behavior. Some
 measurements of cold cores in the Taurus region between 160 and 2100
 $\mic$, do not show departures of $\beta$ from
 $\beta=2$ \citep[see, for instance,][]{Schnee10}. In addition, we
 know that the FIR-to-mm emission varies as a function of
wavelength, as observed in laboratory experiments.
\citep{Boudet05, Coupeaud11}.
\begin{figure}[!t]
\begin{center}
\includegraphics[width=8cm]{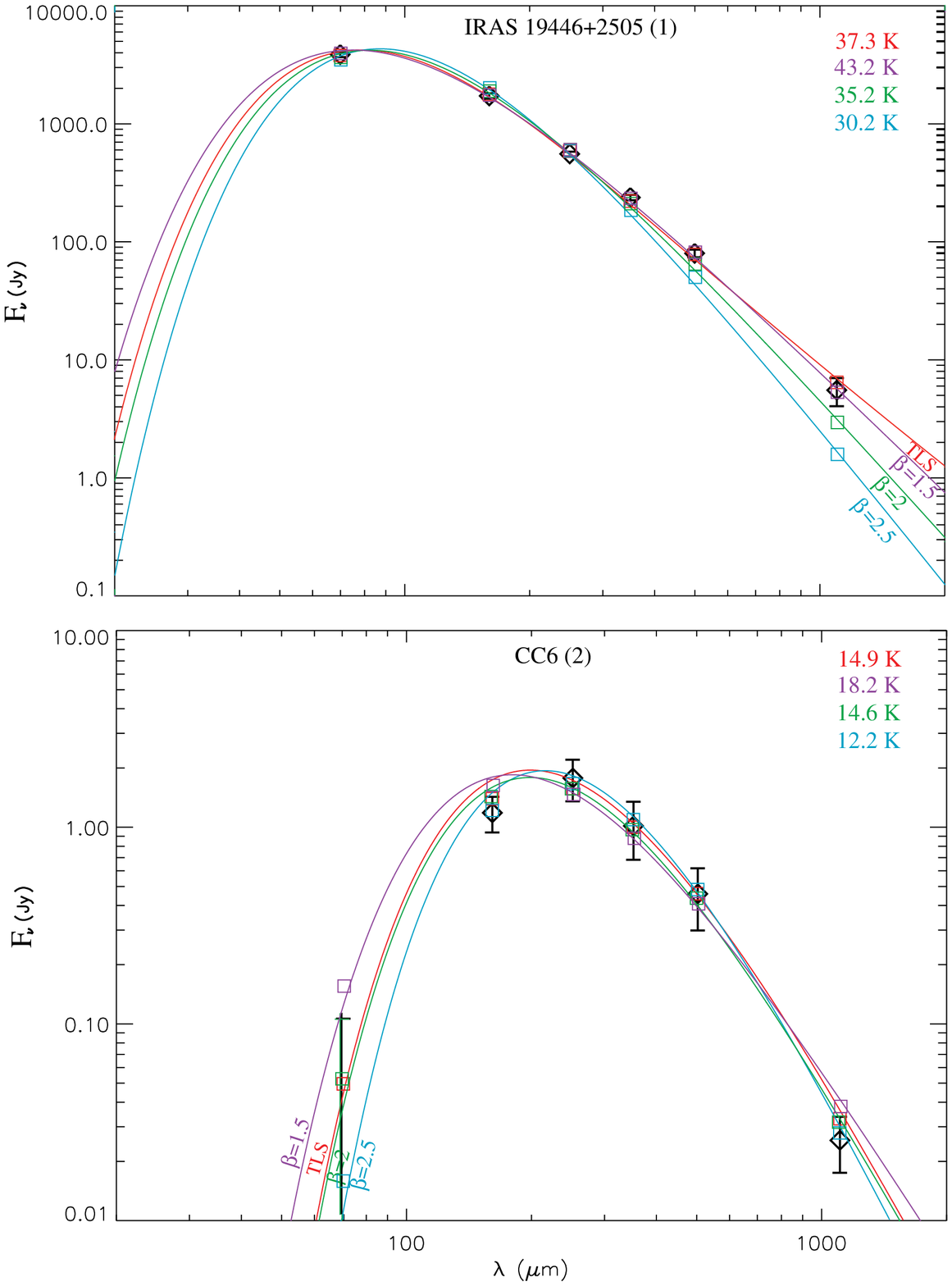}
\caption{Herschel (70, 160, 250, 350, 500 $\mic$) - Bolocam (1.1 mm)
  SED (black diamonds) of IRAS 19446+2505 (1) and CC6 (2), fitted with the TLS model using
  the standard parameters in red, and with a T-$\beta$ model
  $\beta$=1.5 in purple, $\beta$=2 in green, and $\beta$=2.5 in
  blue. Squares represent models integrated in
  the band filters of each instrument, allowing direct comparisons
  with data (diamonds). The 70 $\mic$ Herschel data have not been
  included in the fits for CC6 (2). The corresponding dust
  temperatures for each model are given in the top right corner. \label{fig_spec_comp}}
\end{center}
\end{figure}

\begin{table*}[t]
\begin{center}
\begin{tabular}{  l c  c  cc |c c cc }
\hline
\hline
 & \multicolumn{8}{c }{Dust temperatures (K)} \\
Regions & \multicolumn{4}{c }{TLS} & \multicolumn{4}{c
 }{T-$\beta$} \\
 & Diff. & CS & Std. & 1-$\sigma$ &$\beta$=2 & $\beta$=1.5 & $\beta$=2.5
 & 1-$\sigma$\\
\hline  
IRAS 17279-3350 (1) & 26.13 & 25.81 & 25.85 & 0.17 & 25.29 & 29.17 & 22.70
& 3.26 \\
IRAS 17279-3350 (2) & 30.72 & 31.27 & 30.73 & 0.31  & 28.68 & 35.29
& 24.71 & 5.34 \\
IRAS 17455-2800 (1) & 29.17 & 29.17 & 29.12 & 0.03 & 28.21 & 33.21 &
24.76 & 4.25 \\
IRAS 17455-2800 (2) & 27.21 & 26.76 & 26.75 & 0.26 & 24.70 & 31.72 &
20.20 & 5.81 \\
IRAS 17577-2320 (1) & 31.10 & 31.17 & 30.81 & 0.19 & 30.14 & 35.08 & 26.67
& 4.23 \\
IRAS 17577-2320 (2) & 27.57 & 27.67 & 27.30 & 0.19 & 25.78 & 31.56 &
22.26 & 4.70 \\
IRAS 18032-2032 (1) & 31.67 & 31.80 & 31.66 & 0.08 & 30.29 & 36.22 &
26.61  & 4.85 \\
IRAS 18032-2032 (2) & 31.19 & 33.22 & 31.21 & 1.17 & 27.44 & 36.19 &
22.53 & 6.92 \\
IRAS 18116-1646 (1) & 31.71 & 32.03 & 31.69 & 0.19 & 30.57 & 36.29 &
26.65  & 4.85 \\
IRAS 18116-1646 (2) & 32.63 & 34.25 & 32.68 & 0.92 & 29.13 & 37.68 &
24.59  & 6.65 \\
IRAS 18317-0757 (1) & 36.65 & 37.27 & 36.69 & 0.35 & 34.92 & 42.66 &
30.03 & 6.37 \\
IRAS 18317-0757 (2) & 27.78 & 28.25 & 27.76 & 0.28 & 25.72 & 32.20 &
21.75  & 5.28 \\
IRAS 18434-0242 (1) & 37.76 & 38.76 & 38.19 & 0.50 & 36.19 & 44.21 &
30.75  & 6.77\\
IRAS 18434-0242 (2) & 26.20 & 26.14 & 26.13 & 0.04 & 24.48 & 30.62 &
20.69  & 5.01 \\
IRAS 18469-0132 (1) & 28.14 & 28.15 & 28.11 & 0.02 & 27.21 & 31.69 &
24.16  & 3.79 \\
IRAS 18469-0132 (2) & 33.54 & 34.76 & 33.63 & 0.68 & 30.73 & 38.73 &
26.15  & 6.37 \\
IRAS 18479-0005 (1) & 32.69 & 33.15 & 32.69 & 0.27 & 31.26 & 37.28 & 27.24
& 5.05 \\
IRAS 18479-0005 (2) & 29.26 & 29.75 & 29.24 & 0.29 & 27.03 & 34.20 &
22.72  & 5.80 \\
IRAS 18502+0051 (1) & 28.10 & 28.10 & 28.06 & 0.02 & 27.14 & 31.72 &
23.77  & 3.99 \\
IRAS 18502+0051 (2) & 24.04 & 23.57 & 23.77 & 0.24 & 22.52 & 27.79 &
18.78  & 4.53 \\
IRAS 19442+2427 (1) & 31.19 & 31.26 & 31.18 & 0.04 & 30.17 & 35.25 &
26.27 & 4.50 \\
IRAS 19442+2427 (2) & 28.74 & 29.15 & 28.71 & 0.25 & 26.96 & 33.23&
23.17  & 5.08 \\
IRAS 19446+2505 (1) & 37.20 & 38.23 & 37.27 & 0.58 & 35.24 & 43.23 &
30.20  & 6.57 \\
IRAS 19446+2505 (2) & 37.63 & 41.25 & 38.13 & 1.96 & 33.56 & 43.61 &
27.74  & 8.03 \\
Mean std. deviation &- &-&-& 0.38  &-&-&-& 5.33 \\
 \hline
Cold clump 1 (1)  & 18.65 & 17.96 & 18.52 & 0.37 & 18.03 & 21.56 &
15.66 & 2.97 \\
Cold clump 1 (2)  & 16.63 & 15.62 & 16.38 & 0.53 & 15.96 & 19.94 &
13.88 & 3.08  \\
Cold clump 2 (1)  & 20.40 & 19.60 & 20.17 & 0.41 & 19.58 & 23.78 &
16.93 & 3.45 \\
Cold clump 2 (2)  & 17.01 & 15.58 &  16.59  & 0.74 & 16.21 & 20.64 &
13.80 & 3.47 \\
Cold clump 3 (1)  & 17.11 & 15.76 & 16.86 & 0.72 & 16.44 & 20.17 &
14.05 & 3.08 \\
Cold clump 3 (2)  & 19.15 & 17.14 & 18.66 & 1.05 & 18.12 & 23.25 &
15.01 & 4.16 \\
Cold clump 4 (1)  & 14.61 & 13.97 & 14.47 & 0.34 & 14.17 & 16.50 &
12.64 & 1.94 \\  
Cold clump 4 (2)  & 16.98 & 15.26 & 16.52 & 0.89 & 16.12 & 20.18 &
13.65 & 3.29 \\
Cold clump 5 (1)  & 14.17 & 13.16 & 13.98 & 0.54 & 13.79 & 16.13 &
12.04 & 2.05 \\
Cold clump 5 (2)  & 17.06 & 14.46 &  16.52 & 1.37 & 16.10 & 20.73 &
13.19 & 3.80 \\
Cold clump 6 (1)  & 11.71 & 10.58 & 11.54 & 0.61 & 11.47 & 13.20 &
10.03 & 1.59  \\
Cold clump 6 (2)  & 15.19 & 13.30 & 14.89  & 1.02 & 14.59 & 18.18 &
12.24 & 2.99 \\
Cold clump 7 (1)  & 13.64 & 12.69 & 13.55  & 0.52 & 13.19 & 15.25 &
11.67 & 1.80 \\
Cold clump 7 (2)  & 19.59 & 17.91 & 19.13 & 0.87 & 18.31 & 24.16 &
15.38 & 4.47 \\
Cold clump 8 (1)  & 12.18 & 11.13 & 12.04 & 0.57 & 11.95 & 13.70 &
10.47 & 1.62 \\
Cold clump 8 (2)  & 16.56 & 14.25 & 16.04 & 1.21 & 15.67 & 20.16 &
12.99 & 3.62 \\
Cold clump 9 (1)  & 18.30 & 17.41 & 18.04 & 0.46 & 17.56 & 21.22 &
15.33 & 2.97 \\
Cold clump 9 (2)  & 17.65 & 15.32 & 17.12 & 1.22 & 16.65 & 21.74 &
13.72 & 4.06 \\
Cold clump 10 (1)  & 19.72 & 19.07 & 19.63 & 0.35 & 19.09 & 22.75 &
16.52 & 3.13 \\
Cold clump 10 (2)  & 20.53 & 18.82 & 20.10 & 0.89 & 19.15 & 25.17 &
15.96 & 4.68 \\
Cold clump 11 (1)  & 13.96 & 13.02 & 13.67 & 0.48 & 13.51 & 15.71 &
11.95 & 1.89 \\
Cold clump 11 (2)  & 16.70 & 15.05 & 16.45 & 0.89 & 16.03 & 20.18 &
13.50 & 3.37 \\
Cold clump 12 (1)  & 15.22 & 14.53 & 15.07 & 0.36 & 14.86 & 17.22 &
13.10 & 2.07 \\
Cold clump 12 (2)  & 14.09 & 12.90 & 13.79 & 0.62 & 13.56 & 16.55 &
11.76 & 2.42 \\
Mean std. deviation & -&-&-& 0.71  &-&-&-& 3.00 \\
 
\hline
\end{tabular}
\end{center}
\caption{Dust temperatures (in K)  derived from the TLS and T-$\beta$ models,
  using different sets of parameters (diffuse, compact sources and
  standard parameters) and $\beta$ values (1.5, 2 and 2.5),
  respectively. 1-$\sigma$ correspond to the standard deviation of
  dust temperature derived from the three set of TLS parameters or
  three $\beta$ values. The mean values of the 1-$\sigma$ columns are also given. \label{tab_temp}}
\end{table*}

\subsubsection{TLS model} \label{sec_tls}
The TLS model is the first model that takes the physical aspect of
amorphous dust material into account. 
We do not give a full description of the TLS model here, but refer to \citet{Meny07} for a theoretical overview of the physics of
the model and to \citet{Paradis11, Paradis12} for comparisons of the
TLS model with astrophysical data (FIRAS/WMAP, Archeops and Herschel
data). In previous analyses, we determined the best parameters that
allowed us to reproduce the Galactic diffuse medium (denoted as
diffuse parameters, or Diff.), Galactic compact sources (denoted as
compact sources parameters, or CS) and both environments
(denoted as standard parameters, or Std.). The TLS model is the combines
two distinct processes: the disordered charge distribution (DCD)
part at the grain scale, and the TLS part itself at the atomic scale. The
first effect describes the interaction between the electromagnetic
wave and acoustic oscillations in the disordered charge of the
amorphous material \citep{Vinogradov60, Schlomann64}. This DCD process is
characterized by a correlation lenght ($l_c$), that controls the
inflection point where two asymptotic behaviors occur ($\epsilon \propto
\lambda^{-2}$ and $\epsilon \propto
\lambda^{-4}$). The TLS process takes the interaction of
the electromagnetic wave with the simple distribution of an asymmetric double-well potential into
account \citep{Phillips72, Phillips87, Anderson72}. This TLS process
is characterized by three specific effects that are
temperature-dependent, which is different from the DCD process. One of these TLS effects is represented by the
parameter $c_{\Delta}$ that describes the tunneling states. 
The amplitude of the TLS effects with respect to the DCD process is controled by a
multiplying factor denoted $A$, that is, $I_{tot}=I_{DCD}+A\sum
I_{TLS}$. \\
In the following we therefore use the three sets of
parameters (Diff., CS, Std.), fixed to some specific values of $l_c$,
$c_{\Delta}$, and $A$ (see Tab. \ref{tab_tls_param}) that were derived from
previous analyses \citep{Paradis11}, when performing SED fitting with the TLS model.
\begin{table*}[!t] 
\caption[]{Results of the polynomial fit (see Eq. 4) for the TLS model
  for each environment. \label{tab_surfit}} 
\begin{center}
\begin{tabular}{l lllll}
 \hline
\hline
Environment & $k_{0,i}$ & $k_{1,i}$ & $k_{2,i}$ & $k_{3,i}$ &
$k_{4,i}$ \\
\hline
Diffuse medium& 3.33830 & -5.36544e-3 & 1.65381e-6 & 1.63754e-11 &
-6.88816e-12 \\
 & -2.98801e-4 & 1.49526e-5 & -1.42524e-7 & 1.41845e-12 & 6.72034e-13
 \\
& 8.54787e-6 & -2.17671e-7 & 1.49001e-9 & 1.03925e-11 & -2.48183e-14
\\
& -3.25327e-8 & 1.06390e-9 & 4.20971e-12 & -1.37459e-13 & 2.39831e-16
\\
& 4.55497e-11 & -1.10996e-12 & -3.84120e-14 & 4.36257e-16 &
-6.77351e-19 \\
\hline
Compact sources & 3.34100 & -6.57687e-3 & 1.67055e-5 & -7.42107e-8 &
7.97846e-11 \\
& -1.78243e-3 & 8.93472e-5 & -1.09494e-6 & 3.82566e-9 & -2.98224e-12
\\
& 4.27571e-5 & -1.54310e-6 & 1.94520e-8 & -5.15802e-11 & 2.51914e-14
\\
& -2.53887e-7 & 1.06665e-8 & -1.19930e-10 & 2.44969e-13 & -2.12459e-17
\\
& 5.47140e-10 & -2.43342e-11 & 2.47310e-13 & -3.67727e-16 &
-2.21749e-19 \\
\hline
Standard medium & 3.33042 & -5.49209e-3 & 2.14115e-6 & -2.14173e-9 &
-7.90430e-12\\
& -2.84206e-4 & 1.12572e-5 & -1.11099e-7 & -7.48509e-11 & 9.32649e-13
\\
& 9.23806e-6 & -1.28278e-7 & 1.17408e-9 & 1.19308e-11 & -3.01610e-14 \\
& -3.20750e-8 & 4.97171e-10 & 5.92135e-12 & -1.52179e-13 & 2.83618e-16
\\
& 3.66206e-11 & 1.88461e-13 & -4.33218e-14 & 4.82933e-16 &
-7.97139e-19 \\
\hline
\end{tabular}
\end{center}
\end{table*}

\begin{table*}[!t] 
\caption[]{Gaussian coefficients from Eqs. 4 and 5 for the three environments. \label{tab_gauss}} 
\begin{center}
\begin{tabular}{l llllll}
 \hline
\hline
Environment & $a_0$ & $a_1$ & $a_2$ & $a_3$ & $a_4$ & $a_5$ \\
\hline
Diffuse medium & -0.00050 & 0.07585& 5.36111 & 100.19785 &
13.47090 & 499.96309 \\
Compact sources & -0.00126 & 0.14795 & 4.52068 & 91.28719 & 9.50025 &
451.84591 \\
Standard medium & -0.00062 & 0.09271 & 5.14453 & 90.71464 & 12.36827 &
484.87299 \\
\hline
\end{tabular}
\end{center}
\end{table*}

\subsubsection{$\chi^2$ minimization}
\label{sec_minimization}
We performed $\chi^2$ minimizations on SEDs using both models
(see Tab. \ref{tab_all}). For the
T-$\beta$ model we applied three
values of $\beta$ (1.5, 2, and 2.5). For the TLS model we used
the three sets of parameters defined in the previous section.  
The three $\beta$ values of 1.5, 2, and 2.5 are not arbitrary
  values. At first order, a mean emissivity spectral index in
  the submm domain derived from the TLS model is close to 2, with the
  use of standard parameters in the range $\sim$17-25 K, 1.5 with diffuse
  parameters in the range $\sim$30-40 K, and 2.5 with CS parameters in the
  range $\sim$8-13 K. In that sense, the choice of these three
  $\beta$ values is similar to that of the submm slope derived from the TLS
  model. However, the slope in the FIR in the TLS model is different
  from the slope in the submm and mm because of the DCD process in the FIR and
  TLS processes in the submm and mm. This change of $\beta$ from FIR to
  submm and mm has been
  observed in various environments \citep[see, for instance,
  ][]{Paradis09, Planck14, Gordon14} 
For the UCHII regions
the $\chi^2$ minimizations was made between 70 $\mic$ and 1.1
mm, while for cold clumps the 70 $\mic$ flux was not included
in the fits. For environmental temperatures higher than $\simeq$25 K, the 70 $\mic$
flux  arises by more than 85$\%$ from big grains in equilibrium with the interstellar
radiation field, according to the DustEM model \citep{Compiegne11}. However, in cold environments, the 70 $\mic$ emission
includes a substantial fraction of emission from small grains
that constantly fluctuate in temperature after a photon
absorption/emission. 
We pre-computed the brightness in the Herschel and Bolocam filters by
applying the color correction necessary for each instrument using
both models, for temperatures ranging from 5 to 50 K, sampled every 0.5
K. The $\chi^2$ value was computed for each value of the grid, and we chose 
the value of the dust temperature that minimizes the $\chi^2$. 
To allow interpolating between
individual entries of the table, the best-fit temperature value ($T^{\star}$) was
computed for the ten lowest values of $\chi^2$ as
\begin{equation}
T^{\star}= \frac{ \sum_{i=1}^{10} T_i\times
  \frac{1}{\chi^2_i}}{\sum_{i=1}^{10} \frac{1}{\chi^2_i}}.
\end{equation}
Temperatures derived from the fits are given in Tab.
\ref{tab_temp}. 
The models were adjusted to the data by adopting the following
  normalization:
\begin{equation}
F_{model,norm}(\lambda)=\frac{F_{model}(\lambda) \times \sum_{\lambda}
  F_{obs}(\lambda)}{\sum_{\lambda}F_{model}(\lambda)},
\end{equation} 
where $F_{model}$ and $F_{model,norm}$ are the integrated flux in each
band deduced from the model before and after normalization,
respectively, and $F_{obs}$ is the observed flux. The sum over
the fluxes is performed between 70 and 500 $\mic$ for SEDs of UCHII regions,
and between 100 and 500 $\mic$ for SEDs of cold clumps.
We note that the dispersion in temperature values can
be significant from one model to the other and from one set of
parameters to the other, but in the latter case the $\chi^2$ dispersion is
high as well. For instance, Tab. \ref{tab_temp} shows that the mean
value of temperature dispersion is 5.33 K and 3.00 K for UCHII regions
and cold clumps with
T-$\beta$ models while it is of 0.38 K and 0.71 K for the TLS model. The comparison
of temperatures derived from fits with the TLS  model (diffuse
parameters) and T-$\beta$ model ($\beta=1.5$) with similar $\chi^2$
illustrates the dispersion: 6.45 K for IRAS 18434-0242 (1)
(37.76 K and 44.21 K for the TLS and T-$\beta$ model);
$\sim$5 K for IRAS 18469-0132 (2) (33.54 K and 38.73 K)
and IRAS 18032-2032 (2) (31.19 and 36.19 K). The
dispersion is lower for the fitting of cold clumps (with CS
parameters and $\beta$=2.5): 1.71 K for CC3 (1) (15.76 K and 14.05 K);
1.61 K for CC4 (2) (15.26 K and 13.65 K); and 1.27 K for CC5 (2)
(14.46 K and 13.19 K). The results show
that the choice of the model has a real and strong impact on the temperature
determination. 
In the TLS model,
the temperature determination is much less sensitive to the slope of
the emissivity at long wavelengths. This is because, in
agreement with laboratory data on silicates between 10 K and 100 K, the
slope of the emissivity starts flattening with temperature in the
submm range, while the temperature is mainly deduced from the FIR
domain ($\lambda < 350$ $\mic$) in the observational
data. Indeed, in the framework of the TLS model, an observed dust emissivity
index far from a value equal to 2 in the 100-350 $\mic$ range cannot
arise from intrinsic properties of silicate grains, but only from a possible
grain temperature distribution and from big grains containing carbon,
for instance. On the other hand, in a T-$\beta$ model, the dust emissivity index is kept constant over the
whole FIR-to-mm range and consequently in the range near the peak of emission,
which is required for an accurate temperature determination. Therefore, for futures studies on optical properties
variations with environment (and temperature as a consequence), the
TLS model does not present the same artifact in terms of
  temperature determination as a T-$\beta$ model, and is in particular
  a better description
of the FIR-to-mm emission. \\

\section{Discussion}
\label{sec_discussion}
The different trends observed between the two types of environments
were deduced from the statistics from 12 regions and the analysis of
24 SEDs (since we have two SEDs per region) for each
environment. We essentially focused on the total sum of the
$\chi^2$, and on the number of best fits (best $\chi^2$) depending
on the model and its associated parameters (see Tab. \ref{tab_all}). To assign the same weight to each SED we normalized the
$\chi^2$ values (Tab. \ref{tab_all}), allocating the value of 1 to the
highest $\chi^2$
value derived from the TLS model for each SED. This
normalization ensures that the total $\chi^2$ is not unaffected by a
single high $\chi^2$ value due to a bad fit. We also normalized
the $\chi^2$ for T-$\beta$ modeling and kept the same reference
value. 
 We have checked the consistency of the SED fitting results by
  allowing the flux density measurements to vary within the range
  permitted by their uncertainties. Although the Herschel
  data are internally calibrated, the zero level of the background in
  both PACS and SPIRE data is not. For the Hi-GAL data, a strategy was
  adopted to set this background level using the IRIS and Planck
  calibrations (see Sect. \ref{sec_data}). For the cold clumps, in
  particular, the entire SED wavelength range from 160 to 500 $\mic$
was cross-calibrated using the Planck data. This allowed us
to have consistent flux density uncertainties, so that the spectral
shape was not affected by the above method. Moreover, the
Bolocam data have large intrinsic uncertainties that are dominated by
the noise \citep{Bally10}; they do not affect the SED fits and
therefore were not touched during the tests. For the UCHII
region SEDs, the 70 $\mic$ measurements were cross-calibrated
using a combination of IRIS and Planck calibration, therefore we 
experimented some more with the uncertainties. We performed two
tests; one consisting of shifting the 70 $\mic$ values up according to
the derived uncertainty of the IRIS 70 $\mic$, and shifting the values
from 160 to 500 $\mic$ down, using the Planck uncertainties. We then
perfomed the opposite case (shifting the 70 $\mic$ flux down and
shifting the flux from 160 to 500 $\mic$ up. For all SEDs, the best
models still correspond to those identified in Tab. \ref{tab_all},
that is, we obtained similar results. Only the dust temperatures are
affected by exploring the range allowed by the uncertainties, with
variations of 1 K to 2.5 K depending on the SED fit. 

We recall that temperature mixing was not
taken into account in this analysis. This possible effect
would affect TLS and T-$\beta$ models in the same way by
inducing a flattening of the spectrum at first order. However, temperature
mixing along the line of sight is expected in the inner Galactic
plane. The Galactic Center, which is particularly exposed to these
effects, showed steep spectra \citep{Paradis12}, which contradicts
expectations. Moreover, \citet{Paradis09} investigated
the effect of the interstellar radiation field strengh mixture (as well as
grain size distribution and grain composition) in cold molecular
clouds to explain the steeper emissivity spectra in the FIR than in
  the submm and mm. They concluded that these effects are
responsible for the submm and mm SED flattening. Even if temperature mixing might have an impact on the spectral
behavior of dust emission, it is unlikely that this would
significantly affect the
conclusions of our analysis here. 

\subsection{ Specific dust properties in each environment}

With the TLS modeling, the total number of best fits deduced from
best $\chi^2$ indicates that compact source (CS) parameters do not give the best description of spectra in UCHII
regions (best $\chi^2$ for only 12$\%$ of the SEDs), while diffuse and standard
parameters give better solutions (52$\%$ and 36$\%$). 
For cold clumps, the former set of parameters (CS parameters)
is satisfactory for 56$\%$ of the SEDs (against 8$\%$ and 36$\%$
for diffuse and standard parameters). 
These results cleary show that SEDs from UCHII regions and cold
clumps are not reproduced by the same set of parameters; therefore
they have different dust properties. 

The results are similar for the T-$\beta$ models. Indeed, 62.5$\%$ of UCHII region
SEDs are well reproduced using $\beta$=1.5, 37.5$\%$ using $\beta$=2,
and no SEDs are compatible with $\beta$=2.5. From the total $\chi^2$ value
(15.5 and 24.7 for $\beta$=1.5 and 2), it
appears that the more reasonable value of $\beta$ is 1.5. Conversely, only 4$\%$ of 
the cold clump SEDs have the best $\chi^2$ using a $\beta$ of 1.5. To describe
cold clumps, the number of best $\chi^2$ are equally distributed between $\beta$=2 and
$\beta$=2.5 (48$\%$), and the total $\chi^2$ is similar as well (14.5
and 18.0 for $\beta$=2 and 2.5). This change in $\beta$ (from 1.5 to 2-2.5) documents a
steepening in the long-wavelength SEDs (500 - 1100 $\mic$). However, possible changes in the emission spectral
shape between 160 $\mic$ and 1.1 mm are not taken into account in this
model. Therefore $\beta$ could be higher between 160 and 500 $\mic$
than at the long
wavelength range (500 $\mic$ to 1.1 mm), as already observed in
\citet{Paradis09}, but would not be detected in this analysis. The
opposite behavior (increase of $\beta$ with wavelength) would 
not be visible either. We do not pretend that $\beta$=1.5 and $\beta$=2 or 2.5 are
the best values to fit spectra for each
environment. Slightly different values ($\beta$$\sim$1.6 for UCHII regions and
$\beta$$\sim$2.3 for cold clumps) seem to better fit the
SEDs. But values of $\beta$ equal to 1.5, 2,
and 2.5 at first order agrees with values
derived from the TLS model (see Sect. \ref{sec_minimization}). In
the same way, a better optimized set of
the three TLS parameters could be obtained. This study is beyond the
scope of this paper. In general, the results suggest that $\beta$
changes with the environment. 

Another important result is that the CS parameters used
to reproduce the Archeops compact sources in our Galaxy
\citep[see][]{Paradis11} are also the
best parameters to describe the Galactic cold clumps analyzed in this
work, considering the total number of best $\chi^2$. This result
indicates that the same set of parameters is able to reproduce various
cold sources observed with different instruments at different
wavelengths. This points out that all cold clumps
have similar general properties.   
Fifty-two percent of the SEDs of our UCHII regions can be reproduced
by using the diffuse parameters when fitting with the TLS model. However, the difference with
standard parameters in terms of total $\chi^2$ or number of best
$\chi^2$ is not significant. In the past, the
lack of data characterizing warm environments in the FIR-mm domain did not allow deriving
TLS parameters for these regions. With dust emission SEDs in
UCHII regions, we tried to determine the best TLS parameters using the same method  
as in \citet{Paradis11} when fitting the Archeops cold clumps. We
performed a $\chi^2$ minimization on the 24 SEDs of UCHII regions
with the same set of parameters (to be determined), allowing only temperature variation
from one SED to another. We searched for the best set of parameters
to describe our full sample of UCHII region SEDs. The large
  uncertainties on the SEDs made the $\chi^2$ minimization
  difficult. They had little
  effect on the reduced $\chi^2$ value. For instance, the difference
  between the diffuse and standard parameters in the minimization of UCHII
  region SEDs is small, only 1.27 and 1.28.  We obtained
  a best reduced $\chi^2$ of 1.25 with new parameters for UCHII regions, that is not
  significant. Moreover, the behavior of the model with these
  same parameters as a function of temperature and
  wavelength is similar to that using diffuse parameters. For this
  reason, we did not derive a new set of parameters for UCHII
  regions in this analysis. However, the use of CS parameters to fit UCHII region SEDs significantly
  increases the reduced $\chi^2$ value (2.09), which confirms that UCHII
  regions have different properties from cold clumps.  A summary of the TLS parameters
characterizing various environments derived from this work and
from previous analyses is given in Tab. \ref{tab_tls_param}.
  


\subsection{Comparing TLS and T-$\beta$ models} 

The best total $\chi^2$ for each model (TLS and
T-$\beta$) are almost identical, regardless of the environment. This means that
modeling  with the TLS model using the adequate set of parameters, or a T-$\beta$ model
using the adequate $\beta$, has the same result for the goodness of fit because of the lack of
  strong constraints at long wavelengths that are crucial to determine the
  divergence between the models. Standard parameters in the TLS model adapt
well in all cases (diffuse medium, compact sources, UCHII regions). In terms of total $\chi^2$, standard
parameters are able to reproduce the emission of each type of
environment  well. This is the first model that is able to describe various types of
medium with a single set of parameters reasonably well by only
changing the dust temperature. For a T-$\beta$ model predictions of emission spectra in a
specific environment require $\beta$ to be known. Otherwise,
predicted emission spectra can lead to incorrect descriptions (and poor $\chi^2$) of dust
emission in some regions. 

\begin{figure*}[!t]
\begin{center}
\includegraphics[width=15cm]{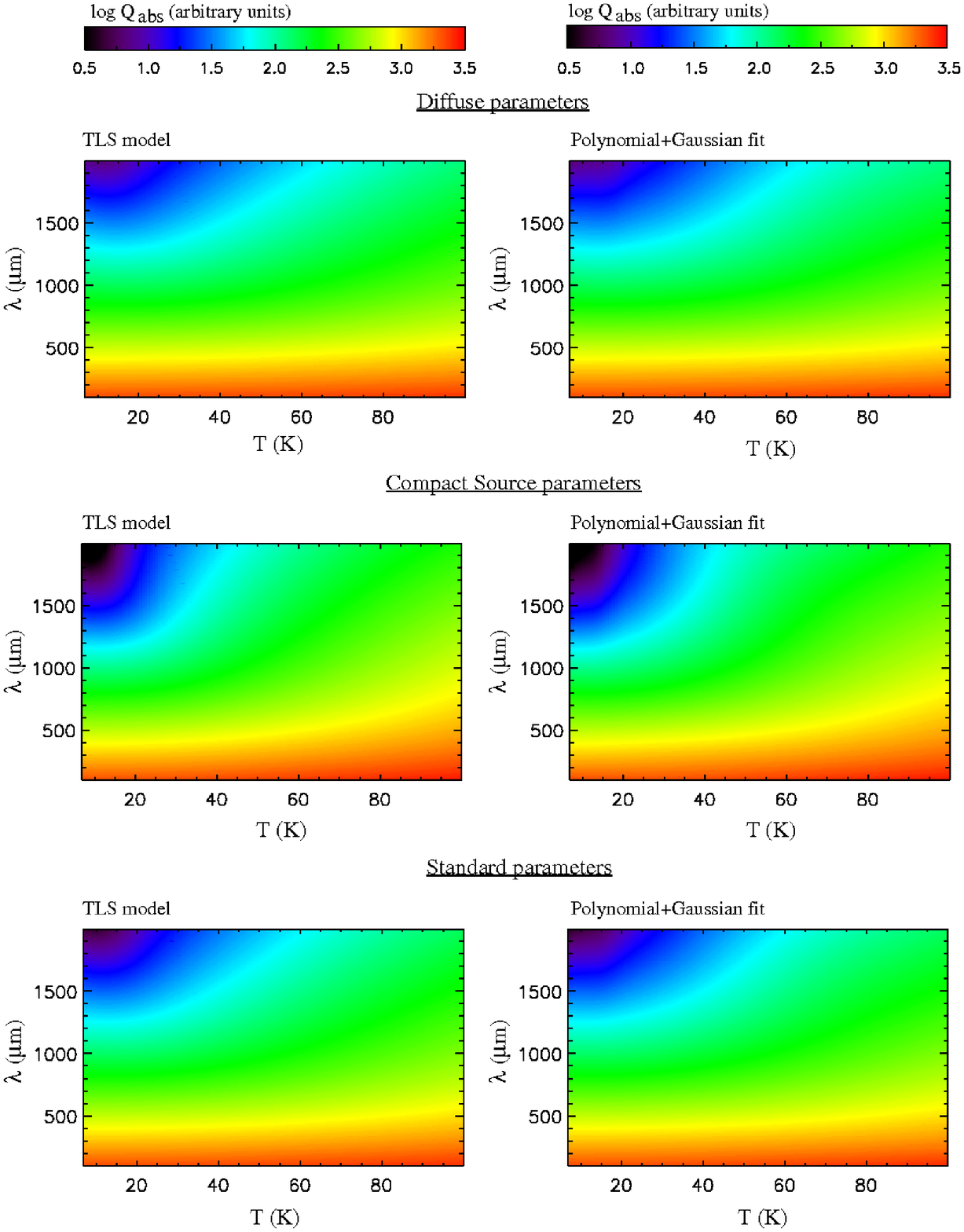}
\caption{ log $\rm Q_{abs}$ (in arbitrary units) as a function of
  temperature and wavelength derived from the TLS
  model (left panels) and from a polynomial+Gaussian fit performed on the TLS
  model (right panels), using different sets of parameters (from top
  to bottom: diffuse, compact sources, and standard).} \label{fig_surf}
\end{center}
\end{figure*}

Figure \ref{fig_spec_comp} shows two SEDs (one for an UCHII region
and one for a cold clump) adjusted with the TLS model (using standard
parameters) and T-$\beta$ models (using $\beta$=1.5; 2; and 2.5). For
IRAS 19446+2505 (1) and CC6 (2), the temperatures reach from 30.2 K to 43.2
K and from 12.2 K to 18.2 K, depending on the
model. Model fluxes after color correction, integrated into each
band filter (squares in the figure), can be directly compared with the
observational SEDs (diamonds in the figure). The
70 $\mic$ flux was not included in the fit for the cold clumps. While a T-$\beta$ model adopting a $\beta$
value of 1.5 is able reproduce the SED of the UCHII region, the same
model gives a poor description of CC6 (2) SED, which requires a steeper
spectrum ($\beta \simeq$2.5). The TLS model, however, describes
each environment quite well by only changing the dust temperature. 
The main difference between the TLS and the T-$\beta$ model with a
reasonable value of $\beta$ occurs in the mid-FIR ($\lambda
<$70$\mic$) and in the mm range. But because of the large
uncertainties in the Bolocam data, especially for cold clumps where
the flux can be at the same level as the noise, the 1.1 mm flux does not add any
strong constraints. However, even though in most cold clumps removing
these data from the fits gives similar results, the 1.1 mm flux can also help the fit in some cases. Moreover, the
1.1 mm flux appears to have the same rough estimate as expected,
which makes us confident in the use of these data.

In summary, each environment is characterized by a different
dust emissivity index of the dust emission, which indicates distinct dust
properties that leads to a change in $\beta$ for a T-$\beta$ model (from $\simeq$2-2.5 to 1.5, corresponding to warm
and cold regions), or to a change in the TLS parameters (standard,
diffuse, or CS parameters), for accurate descriptions of each type of environment.
However,  different from a T-$\beta$ model with a fixed $\beta$ that is
not able to give good fits in warm as well as in cold regions, the
standard TLS parameters can reproduce all types of environment
reasonably well. 

\subsection{Simplifications of dust emission modeling} 
\subsubsection{Polynomial fit on the TLS model} 
\label{sec_polyfit}
To facilitate using the TLS model predictions as a function of the
environment, we performed a polynomial fit on the model, using the
idl function {\it sfit}, for 
each set of parameters (diffuse, cold sources, and standard). This idl function allows us to determine a polynomial fit to a
surface, which in our case is the dust absorption efficiency
($Q_{abs}$) deduced from the model as a
function of temperature (6.9 K - 100 K) and wavelength (100 $\mic$ - 2
mm). However, for temperatures between 10 K and 15 K at
  wavelengths of around 2 mm, the difference between the
  model and the polynomial fit might become important. To minimize the
difference we included a Gaussian function in the fit (using the idl
function {\it gauss2dfit}). The final 2D function\footnote{IDL
  code available
  here: http://userpages.irap.omp.eu/$\sim$dparadis/TLS/
compute$\_$TLS$\_$poly$\_$gaussian$\_$fit.pro} used to fit the
model is then given as follows:
\begin{eqnarray} \label{eq_sfit}
&& log Q_{abs}\left ( \lambda, T\right )  =  \sum k_{j,i} \left ( (log
  \lambda-2.00076)/2.70167 \times 10^{-3} \right )^i  \nonumber \\
&& \times \left ( (T-4.30000)/6.46369\times 10^{-1}\right ) ^j + a_0+
a_1 \times exp (\frac{-U}{2}), 
\end{eqnarray}
with
\begin{equation}
U=\left ( \frac{T-a_4}{a_2} \right )^2 + \left (
  \frac{\lambda-a_5}{a_3} \right ) ^2.
\end{equation}
\begin{figure}[!t]
\begin{center}
\includegraphics[width=8.5cm]{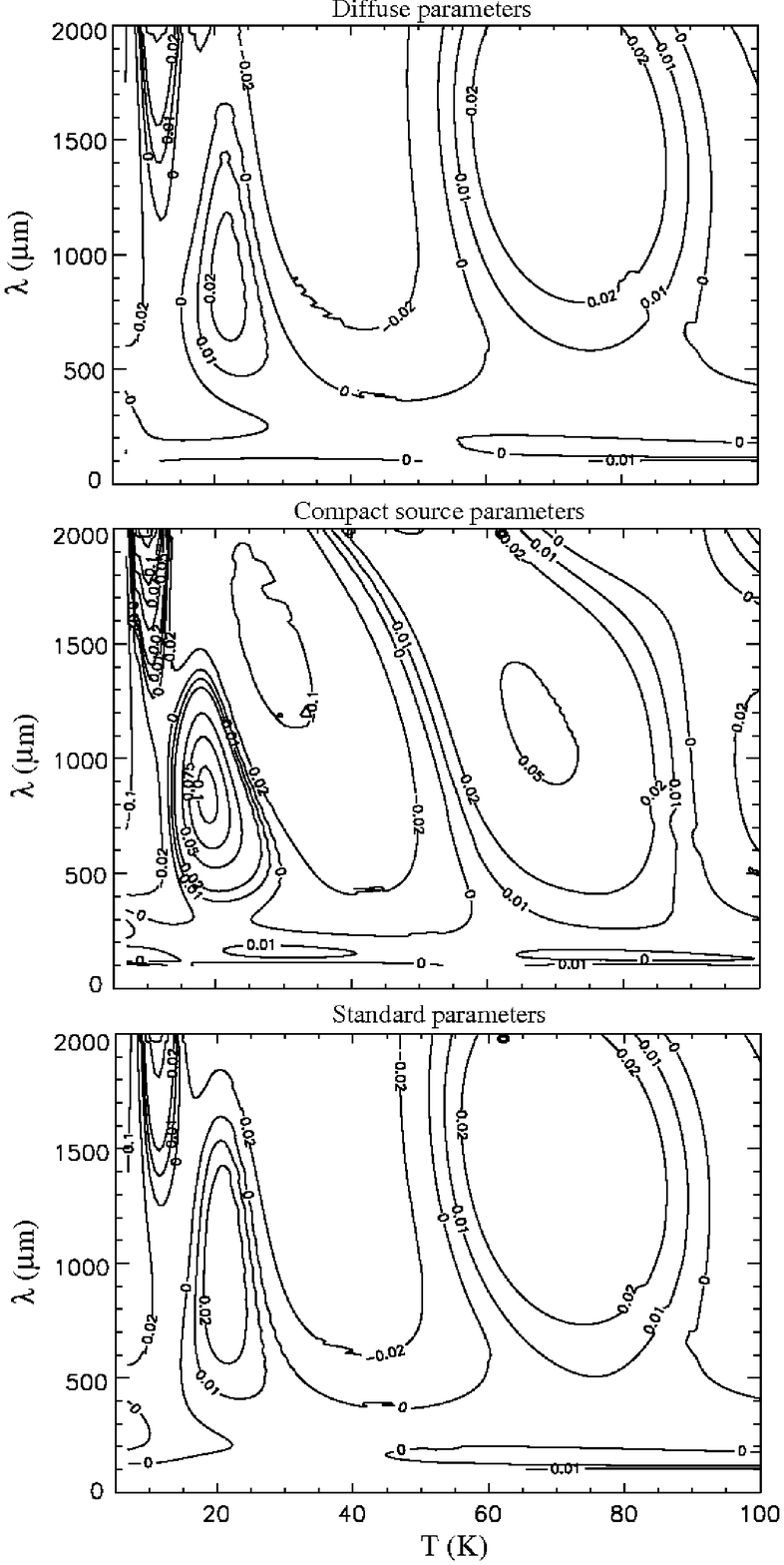}
\caption{Relative error (contours) between the polynomial+Gaussian fit and
  the TLS model $\left ( (Q_{abs,fit}-Q_{abs,model})/Q_{abs,model} \right )$ as a
  function of temperature and wavelength for the three sets of TLS parameters.\label{fig_surf_err}}
\end{center}
\end{figure}

\begin{figure}[!t]
\begin{center}
\includegraphics[width=8.5cm]{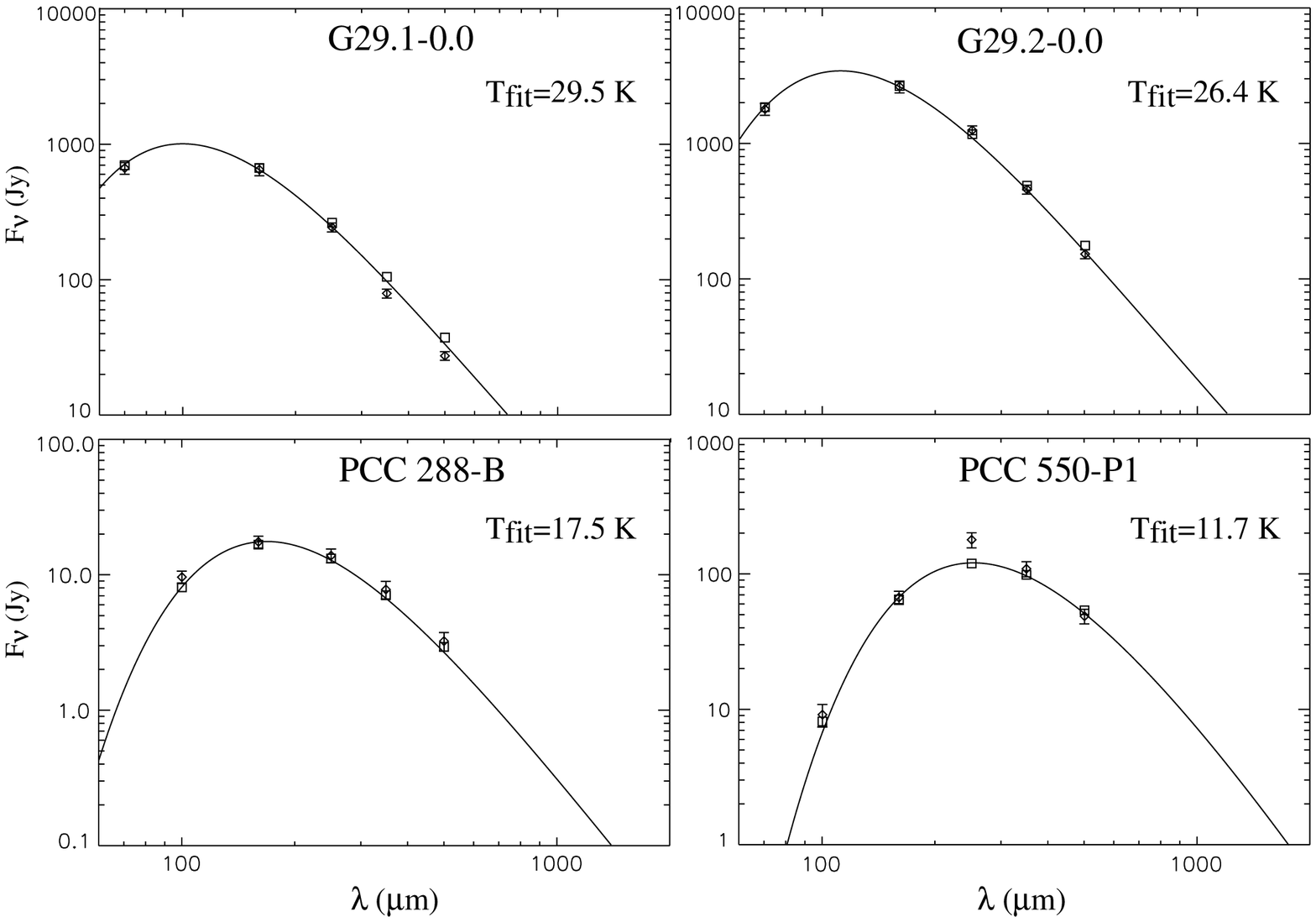}
\caption{Herschel SEDs (diamonds) of two UCHII regions G29.1-0.0 and G29.2-0.0
  from \citet{Paladini12} and two cold cores PCC 288-B and PCC 550-P1
  from \citet{Juvela10}. Polynomial+Gaussian fits of the TLS model are
  represented by the continuous line between 100 $\mic$ and 2
  mm. Squares represent models integrated in
  the band filters of each instrument which allows direct comparisons
  with the data (diamonds). \label{fig_poly}}
\end{center}
\end{figure}
The $k_{j,i}$ and $a$ coefficients are given in Tab.
\ref{tab_surfit} and \ref{tab_gauss}. The
wavelength range is limited to 100 $\mic$ in the fits because for cold
environments, emission at wavelengths below this limit can be
contaminated by emission from small grains that are not in equilibrium
with the radiation field. But since the polynomial fit is linear
  with wavelength in the FIR, that is, for $\lambda < 300 \mic$ in logarithmic
  scale, it can easily be extrapolated to shorter wavelengths if necessary.
We found that a degree of 4 is adequate to achieve a reasonable fit on
the TLS model. Plots of the TLS model and polynomial+Gaussian fits are
presented in Fig. \ref{fig_surf}. Predictions of dust emission
derived from the TLS model (proportional to $Q_{abs}$ values) as well as from the polynomial+Gaussian fit are
given in arbitrary units, which means that they have to be normalized
before they can be used. Reference values of
emissivity or optical constants are given in the
litterature, for instance, \citet{Boulanger96} and \citet{Li01}, which can then
be converted into dust absorption efficiency. The normalization of the
  surface ($Q_{abs}$) to a reference value at some wavelength and
  temperature would lead to an easy determination of the dust
  column density of any observations in the framework of the TLS
  model. However, reference values of emissivity were determined
  for a given wavelength and for a specific temperature. The TLS model
  predicts emissivity variations as a function of wavelength and
  temperature. Predicted emissivities in the IRAS, Herschel, and Planck
bands are given in \citet{Paradis11} for different temperatures
between 5 K and 100 K. 
Contours of the relative error
$ \left ( (Q_{abs,fit}-Q_{abs,model})/Q_{abs,model} \right )$ for
each set of TLS parameters are given in Fig.
\ref{fig_surf_err}. The 1-$\sigma$ standard deviation on the relative error is 3$\%$ over the entire ranges of temperatures and
wavelengths for the CS parameters, and less than 2$\%$ for the diffuse
and standard parameters. For all sets of
  parameters, absolute errors can also reach 10-12$\%$ at low temperature ($\sim
  7$ K) in the submm and/or mm domain. We therefore encourage 
  considering only temperatures higher than 7.5 K  when using the
  polynomial+Gaussian fit. In addition, one has to be careful when
  using CS parameters: we note an increase in the
  errors when reaching long wavelengths (1750-2000 $\mic$),
  for temperatures around 10-15 K and an error of 10$\%$ for
  wavelengths between 700 $\mic$ and 950 $\mic$ and temperatures in the range 17 K - 20 K. 

\subsubsection{Universal application of the polynomial fit}
The interest of the polynomial+Gaussian fit is to describe dust emission SEDs
between 100 $\mic$ (or at shorter wavelengths by extrapolation when
  analyzing warm/hot dust grains) and 2 mm in any regions of our Galaxy. If 
the equilibrium dust temperature is known, it is easy to deduce the
SED. In the opposite case, when the dust emission SED is known, it is
then possible to determine the dust temperature. To check the
applicability of this polynomial+Gaussian fit (performed on the TLS model with
the use of the standard parameters), we compared the fit with known
SEDs of UCHII regions \citep[G29.1-0.0 and G29.2-0.0,
from][]{Paladini12} and cold clumps \citep[PCC 288-B and PCC 550-P1,
from][]{Juvela10}. Only the dust temperature in our fits varied from one SED to
the other. Results are presented in Fig. \ref{fig_poly}. As we showed in Sect. \ref{sec_minimization}, the
determination of the dust temperature depends on the model
used. Here, we did not perform any $\chi^2$ minimization. For
G29.1-0.0 and G29.2-0.0, we used temperature values of 29.5 K and 26.4
K, which is close to the value of the cold
component (29.5 K and 25.6 K) 
derived by \citet{Paladini12} when using a two-component model with
fixed dust emissivity index to
minimize SEDs between 24 and 500 $\mic$. In these two regions the 70
$\mic$ emission is largely dominated by emission from the cold
component. The polynomial+Gaussian fits were performed between 100 $\mic$
and 2 mm (see Sect. \ref{sec_polyfit}) and were extrapolated
to 70 $\mic$ here, as shown in
Fig. \ref{fig_poly}. For the cold clump PCC 550-P1, we considered a
temperature of 11.7 K , which is close to the value of 11.3 K derived by \citet{Juvela10} using a T-$\beta$
model with  
a deduced $\beta$ equal to 2.03. For PCC 288-B, the comparison between
the polynomial+Gaussian fit and the SED is unsuitable when using the dust temperature derived
from \citet{Juvela10} (20.2 K), with a $\beta$ value found equal to 1.36. With
the polynomial+Gaussian fit, a most appropriate value of dust temperature is
around 17.5 K. We recall that the fits presented in Fig.  
\ref{fig_poly} might be even better with the use of CC
parameters in the polynomial+Gaussian fits for the cold clumps with the appropriate dust
temperature. For PCC 550-P1, the fit is not able to reproduce 
the 250 $\mic$ flux, which could be due to calibration problems that
have been improved since the first Herschel data. As reported in
\citet{Juvela10}, a T-$\beta$ model is not able to match the 250
$\mic$ flux either. 

We recall that model predictions essentially differ in the
FIR and long wavelengths and also lead to different dust
temperatures. For this reason, we encourage using the
polynomial+Gaussian fit (or the TLS model), which does not bias the temperature
estimate, but also takes the flattening of the spectra in the
submm-mm domain into account, contrary to T-$\beta$ models. 
  The TLS model predicts a more correct emissivity spectral behavior
  than any single fixed value of $\beta$ and 
  precisely describes the emissivity spectral index as a function
  of temperature and wavelength \citep[see][Fig. 6]{Paradis11}.

\section{Conclusions}
\label{sec_cl}

Using a combination of Herschel and Bolocam Galactic Plane surveys
(Hi-GAL and BGPS) smoothed to a common resolution of 37$^{\prime \prime}$, we
analyzed dust emission associated with two specific environments: UCHII
regions and cold clumps. We studied twelve regions for each
environment. We extracted SEDs in the central and the
surrounding part of each region. We were able to compare the recent
TLS model with emission spectra from warm dust ($\sim 30-40 K$) in
UCHII regions. 
We observed some variations in the dust optical properties
with environments, as revealed by the change in the dust emissivity index, or
in the set of TLS parameters that best fit the emission. In addition, contrary to any fixed value of the dust emissivity index (1.5, 2 and
2.5) that mostly fails to
give good normalized $\chi^2$ in both warm environments such as UCHII
regions and cold clump regions, the use of the standard TLS
parameters can give reasonable results in all cases. These standard
parameters were derived in a previous analysis to reproduce
compact sources observed with Archeops and the diffuse medium as observed
with FIRAS. Using
a T-$\beta$ model for which the $\beta$ value is unknown can
lead to an incorrect description of the dust emission. This comparison shows
that the TLS model can easily be used to reliably predict dust emission spectra
in any region of our Galaxy, in contrast to the T-$\beta$ model.
We also reported an easy way to determine the emission at any
temperature (in the range 7.5 K - 100 K) and wavelength (in the range
100 $\mic$ - 2 mm) for each set of
TLS parameters by giving the 25 coefficients of a polynomial fit
of degree 4, coupled with a Gaussian fit, which accurately reproduces
the BG emission, after it is
normalized to any reference value. The IDL code for the
polynomial+Gaussian fit is available online. 

\begin{acknowledgements}
This research has made use of the NASA/ IPAC Infrared Science Archive,
which is operated by the Jet Propulsion Laboratory, California
Institute of Technology, under contract with the National Aeronautics
and Space Administration. The authors acknowledge the support of the
French Agence National de la Recherche (ANR) through the programme
``CIMMES'' (ANR-11-BS56-0029). Herschel is an ESA space
  observatory with science instruments provided by European-led
  Principal Investigator consortia and with important participation
  from NASA.
\end{acknowledgements}

\evensidemargin=-280pt

\begin{sidewaystable*}[p]
\begin{center}
\begin{tabular}{   l c  c  c c c c |c c c c c c }

\hline
\hline
 Regions & \multicolumn{3}{c }{$\chi^2_{TLS}$} & \multicolumn{3}{c
 }{Norm. $\chi^2_{TLS}$} & \multicolumn{3}{c }{$\chi^2_{T-\beta}$}
 &\multicolumn{3}{c
 }{Norm. $\chi^2_{T-\beta}$} \\
& Diff. &  CS & Std. &Diff. &  CS & Std. & $\beta=2$ & $\beta=1.5$ &
$\beta=2.5$ & $\beta=2$& $\beta=1.5$ & $\beta=2.5$\\
\hline
IRAS 17279-3350 (1) & 0.801 & 0.974 & 0.805 & 0.822 & 1.000 & 0.826 &
 1.074 & 0.426 & 2.351 & 1.103 & 0.437 & 2.414 \\
 IRAS 17279-3350 (2) & 0.099 & 0.250 & 0.112  & 0.396  & 1.000 & 0.448  &
 0.086 & 0.131 & 0.305 & 0.344 & 0.524 & 1.220 \\
IRAS 17455-2800 (1) & 0.061 & 0.112 & 0.062  & 0.545  & 1.000 & 0.554 &
0.211 & 0.166 & 1.170 & 1.884 & 1.482 & 10.446 \\
IRAS 17455-2800 (2) & 0.842 & 1.158  & 0.847  & 0.727 & 1.000 & 0.731 &
0.477 & 0.886 & 0.581 & 0.412 & 0.765 & 0.502 \\
IRAS 17577-2320 (1) & 0.804 & 0.745 & 0.775  & 1.000 & 0.927 & 0.964 &
1.327 & 0.341 &  2.814 & 1.650 & 0.424 & 3.500\\
IRAS 17577-2320 (2) & 0.172 & 0.266 & 0.172 & 0.645 & 1.000 & 0.645 &
0.315 & 0.111 & 0.894 & 1.184 &  0.417 & 3.361 \\
IRAS 18032-2032 (1) & 1.237 & 1.377  & 1.177 & 0.898 & 1.000 & 0.855 &
2.106 & 1.008 & 4.421 & 1.529 & 0.732 & 3.211  \\
IRAS 18032-2032 (2) & 0.408 & 0.911  & 0.448 & 0.448 & 1.000 & 0.492 &
0.049 & 0.373 & 0.191 & 0.054 & 0.409 & 0.210 \\
IRAS 18116-1646 (1) & 0.233 & 0.343 & 0.217  & 0.679 & 1.000 & 0.633 &
0.686 & 0.065 & 2.480 & 2.000 & 0.190 & 7.230 \\
IRAS 18116-1646 (2) & 0.336 & 0.956  & 0.389 & 0.351 & 1.000 & 0.407 &
0.059 & 0.294 & 0.378 & 0.062 & 0.308 & 0.395\\
IRAS 18317-0757 (1) & 0.234 & 0.561  & 0.290 & 0.417 & 1.000 & 0.517 &
0.088 & 0.485 &  0.397 & 0.157 & 0.865& 0.708 \\
IRAS 18317-0757 (2) & 0.288 & 0.443 & 0.291 & 0.650 & 1.000 & 0.657 &
0.296 & 0.239 & 0.741 & 0.668 & 0.539 & 1.673 \\
IRAS 18434-0242 (1) & 4.016 & 3.966 & 3.922  & 1.000  & 0.988 & 0.977 &
5.029 & 4.032 & 6.590 & 1.252 & 1.004 & 1.641\\
IRAS 18434-0242 (2) & 0.069 & 0.131 & 0.068 & 0.527  & 1.000 & 0.519 &
0.075 & 0.130 & 0.297 & 0.573  & 0.992  & 2.267 \\
IRAS 18469-0132 (1) & 1.217 & 1.407 & 1.204  & 0.865 & 1.000 & 0.858 &
1.833 & 0.519 & 4.200 & 1.303 & 0.369 & 2.985 \\
IRAS 18469-0132 (2) & 0.041 & 0.118 & 0.048 & 0.347 & 1.000 & 0.407 &
0.100 & 0.066 & 0.250 & 0.847 & 0.559 & 2.119  \\
IRAS 18479-0005 (1) & 2.087 & 1.968  & 1.983 & 1.000 & 0.943 & 0.950 &
3.345 & 1.578 & 6.143 & 1.603 & 0.756 & 2.943 \\
IRAS 18479-0005 (2) & 0.371 & 0.700 & 0.396 &  0.530 & 1.000 & 0.566 &
0.082 &  0.443 &  0.165  & 0.117 & 0.633 & 0.236 \\
IRAS 18502+0051 (1) & 1.160 & 1.498 & 1.158  & 0.774  & 1.000 & 0.773 &
1.576 & 0.642 & 3.988 & 1.052 &  0.429 & 2.662 \\
IRAS 18502+0051 (2) & 0.052 & 0.073 & 0.050  & 0.712  & 1.000 & 0.684 &
0.096 &  0.136 & 0.483 & 1.315 & 1.863  & 6.616 \\
IRAS 19442+2427 (1) & 2.134 & 2.046 & 2.061 & 1.000 & 0.959 & 0.966 &
3.285 & 1.117 & 6.391 & 1.539 & 0.523  & 2.995 \\
IRAS 19442+2427 (2) & 0.131 & 0.302 & 0.143  & 0.434 & 1.000 & 0.474 &
0.039 &  0.224 & 0.231 & 0.129 & 0.742 & 0.765 \\
IRAS 19446+2505 (1) & 0.249 & 0.301 & 0.211  & 0.827 & 1.000 & 0.701 &
1.121 & 0.123 & 3.275 & 3.724 & 0.409  & 10.880 \\
IRAS 19446+2505 (2) & 0.111 & 0.512 & 0.151 & 0.217 & 1.000 & 0.295 &
0.091 & 0.055 & 0.386 & 0.178 & 0.107 & 0.754 \\
Total nb. of best $\chi^2$ & -& - & -
& 13 (52$\%$) & 3 (12$\%$)& 9 (36$\%$)& -&- & -& 9 (37.5$\%)$&
15 (62.5$\%$) & 0 (0$\%$) \\ 
Total $\chi^2$  &- & -& - & 15.811 & 23.817 & 15.899 &
-& - & - &  24.679 &15.478 & 71.733 \\ 
\hline
Cold clump 1 (1) & 0.017 & 0.021  & 0.017  & 0.810 & 1.000
& 0.810 & 0.017 & 0.027 & 0.026 & 0.810 &  1.286 & 1.238 \\
Cold clump 1 (2) & 0.002 & 0.068  & 0.008 & 0.029 & 1.000
& 0.117 & 0.011 & 0.030 & 0.088 & 0.162 & 0.441 & 1.294 \\
Cold clump 2 (1) & 0.019 & 0.036 & 0.015  & 0.528 & 1.000
& 0.417 & 0.021 & 0.047 & 0.101 & 0.583 & 1.306 & 2.806 \\
Cold clump 2 (2) & 0.037 & 0.104 & 0.031 & 0.356 & 1.000
& 0.298 & 0.030 & 0.115 & 0.121 & 0.288 & 1.106 & 1.163 \\
Cold clump 3 (1) & 0.014 & 0.001 & 0.009  & 1.000 & 0.071
& 0.643 & 0.008 & 0.045 & 0.001 & 0.571 & 3.214 & 0.071 \\
Cold clump 3 (2) & 0.551 & 0.170 & 0.441 & 1.000 & 0.301
& 0.800 & 0.305 & 1.045 & 0.062 & 0.553 & 1.897 & 0.113 \\
Cold clump 4 (1) & 0.007 & 0.023 & 0.006 & 0.304 & 1.000
& 0.261 & 0.008 & 0.029 & 0.023 & 0.348 & 1.261 & 1.000\\
Cold clump 4 (2) & 0.255 & 0.022 & 0.185 & 1.000 & 0.086
& 0.725 & 0.147 & 0.701 & 0.021 & 0.576 & 2.749 &  0.082 \\
Cold clump 5 (1) & 0.028 & 0.003 & 0.019 & 1.000 & 0.107
& 0.679 & 0.020 & 0.090 & 0.003 & 0.714 & 3.214 & 0.107 \\
Cold clump 5 (2) & 0.270 & 0.026 & 0.200 & 1.000 & 0.096
& 0.741 & 0.188 & 0.556 & 0.023 & 0.696 & 2.059 & 0.085 \\
Cold clump 6 (1) & 0.243  & 0.056 & 0.200 & 1.000 & 0.230 &
0.823 & 0.216 & 0.427 & 0.098 & 0.889 & 1.757 & 0.403 \\
Cold clump 6 (2) & 0.169  & 0.015 & 0.114 & 1.000 & 0.089 &
0.675 & 0.121 & 0.422 & 0.014 & 0.716 & 2.497 & 0.083 \\
Cold clump 7 (1) & 0.099  & 0.119 &  0.097 & 0.832 & 1.000
& 0.815 & 0.103 & 0.103 & 0.128 & 0.866 & 0.866  &  1.076 \\
Cold clump 7 (2) &  0.121 & 0.025 & 0.076 & 1.000 & 0.207 &
0.628 & 0.030 & 0.438 & 0.188 & 0.248 & 3.620 & 1.554 \\
Cold clump 8 (1) & 0.066 & 0.027 & 0.058 & 1.000 & 0.409
& 0.879 & 0.058 & 0.106 & 0.033 & 0.879 & 1.606  &  0.500 \\
Cold clump 8 (2) & 0.932 & 0.055 & 0.679 & 1.000 & 0.059
& 0.729 & 0.639 & 2.028 & 0.080 & 0.686 &  2.176 & 0.086 \\
Cold clump 9 (1) & 0.049 & 0.102 & 0.033 & 0.480 & 1.000 &
0.324 & 0.033 & 0.166 &  0.184 & 0.323 & 1.627 & 1.804 \\
Cold clump 9 (2) & 0.351 & 0.100 & 0.266 & 1.000 & 0.285 &
0.758 & 0.238 & 0.658 & 0.100  & 0.678 & 1.875 & 0.285 \\
Cold clump 10 (1) & 0.042 & 0.040 & 0.042  & 1.000 & 0.952 &
1.000 & 0.042 & 0.052 & 0.043  & 1.000 & 1.238 & 1.024 \\
Cold clump 10 (2) & 0.254 & 0.071 & 0.189 & 1.000 & 0.280
& 0.744 & 0.053 & 0.604 & 0.088 &  0.209 & 2.378 & 0.346 \\
Cold clump 11 (1) & 0.010 & 0.008 & 0.007 & 1.000 & 0.800
& 0.700 & 0.009 & 0.018 & 0.009 & 0.900 & 1.800 & 0.900  \\
Cold clump 11 (2) & 0.264 & 0.099 & 0.203 & 1.000 & 0.375 &
0.769 & 0.189 & 0.595 & 0.095 & 0.716 & 2.254 & 0.360 \\
Cold clump 12 (1)  & 0.064  & 0.094 & 0.056 & 0.681 & 1.000
& 0.596 & 0.065 & 0.111 & 0.103 & 0.691 & 1.181  & 1.100 \\
Cold clump 12 (2)  & 0.039 & 0.055 & 0.024 & 0.710 & 1.000&
0.436 & 0.022 &  0.180 & 0.027 & 0.400 & 3.273 & 0.490 \\
Total nb. of best $\chi^2$ & - & - & - & 2 (8$\%$) & 14 (56$\%$) & 9
(36$\%$) & - &- &- & 12 (48$\%$) & 1 (4$\%$) & 12 (48$\%$) \\
Total $\chi^2$ & - & - & - & 19.730 & 13.347 & 15.367 & - & - & - &
14.502 & 46.681 & 17.970 \\
\hline
\end{tabular}
\end{center}
\caption{Reduced $\chi^2$ derived from fits between Herschel-Bolocam
  data and TLS and T-$\beta$ modeling using different sets of
  parameters (diffuse, compact sources, and standard parameters,
  columns 2, 3, and 4, respectively) and
  $\beta$ values (2, 1.5, and 2.5, columns 8, 9, and 10). Values of
  reduced $\chi^2$ presented in columns 5, 6, 7 and 11, 12, 13 are
  normalized with respect to the highest $\chi^2$
value derived from the TLS model. \label{tab_all}}
\end{sidewaystable*}

 
\twocolumn
\oddsidemargin=0pt
\evensidemargin=0pt

\end{document}